\documentclass[fleqn]{jfm}
\ifCUPmtlplainloaded \else
  \checkfont{eurm10}
  \iffontfound
    \IfFileExists{upmath.sty}
      {\typeout{^^JFound AMS Euler Roman fonts on the system,
                   using the 'upmath' package.^^J}%
       \usepackage{upmath}}
      {\typeout{^^JFound AMS Euler Roman fonts on the system, but you
                   dont seem to have the}%
       \typeout{'upmath' package installed. JFM.cls can take advantage
                 of these fonts,^^Jif you use 'upmath' package.^^J}%
      }
  \else
  \fi
\fi
\ifCUPmtlplainloaded \else
  \checkfont{msam10}
  \iffontfound
    \IfFileExists{amssymb.sty}
      {\typeout{^^JFound AMS Symbol fonts on the system, using the
                'amssymb' package.^^J}%
       \usepackage{amssymb}%

      }{}
  \fi
\fi
\ifCUPmtlplainloaded \else
  \IfFileExists{amsbsy.sty}
    {\typeout{^^JFound the 'amsbsy' package on the system, using it.^^J}%
     \usepackage{amsbsy}}
    {}
\fi

\newsavebox{\astrutbox}
\sbox{\astrutbox}{\rule[-5pt]{0pt}{20pt}}

\newcommand\etal{\mbox{\textit{et al.}}}

\newcounter{saveeqn}
\usepackage{epsfig}
\usepackage{psfrag}
\usepackage{amsfonts}          
\usepackage{amssymb}           
\usepackage{amsmath}           
\usepackage{hhline}            

\renewcommand{\thesection}{\arabic{section}}

\newcounter{eee}

\newcounter{eeeb}

\newcommand{\mypsfrag}[2]{\psfrag{#1}{\footnotesize{#2}}}
\newcommand{\npsfrag}[3]{\psfrag{#1}[#2]{\footnotesize{#3}}}

\newcommand{\x}{\cdot}

\newcommand{\dd}{\partial}
\newcommand{\de}{{\rm \, d}}

\renewcommand{\vec}[1]{\mbox{\boldmath $ #1$}}

\newcommand{\bec}{\vec}

\renewcommand{\P}{$P\ $}

\newcommand\ie{i.e.\ }


\usepackage{color}

\newcommand{\myfigure}[3]{
\begin{figure}
#1
\caption[]{#2}
\label{#3}
\end{figure}
}

\title[Parameter dependences of convective dynamos]{Parameter dependences of
  convection-driven dynamos in rotating spherical fluid shells}

\author[F. H. Busse and R. Simitev]%
{FRIEDRICH H.~BUSSE$\ast$\thanks{Email: \tt busse@uni-bayreuth.de} and RADOSTIN D.~SIMITEV$\ast$\ddag}
\affiliation{
$\ast$~Institute of Physics, University of Bayreuth, D-95440  Bayreuth, Germany\\
\ddag~Department of Mathematical Sciences, The University of Liverpool, Liverpool L69 7ZL, UK\\
}
\date{\today}

\begin{document}

\maketitle

\begin{abstract}\centering\parbox{12cm}{
For the understanding of planetary and stellar dynamos an overview of
the major parameter dependences of convection driven dynamos 
in rotating spherical fluid shells is desirable. Although the
computationally accessible parameter space is limited, earlier work is
extended with emphasis on higher Prandtl numbers and uniform heat flux
condition at the outer boundary. The transition from dynamos dominated
by non-axisymmetric components of the magnetic field to those dominated by the
axisymmetric components depends on the magnetic Prandtl number as well
as on the ordinary Prandtl number for higher values of the rotation
parameter $\tau$. The dependence of the transition on the latter
parameter is also discussed. A variety of oscillating dynamos is
presented and interpreted in terms of dynamo waves, standing
oscillation or modified relaxation oscillations. 
}\end{abstract}

\section{Introduction}

Most global magnetic fields of planets and stars are generated by
thermal or compositional convection in the deep interiors of those
bodies. In spite of their common origin a large variety of dynamo
processes is indicated by the observations of planetary and stellar
magnetic fields. This is not surprising in view of the widely varying
conditions  under which dynamos operate. It is thus desirable to
understand the effects of the most influential parameters on
convection driven dynamos. The purpose of a study of the parameter
dependence of dynamos would be twofold: On the one hand properties of
observed magnetic fields could be explained in terms of parameter
values of the respective system. On the other hand, unknown conditions
in the interior of the celestial bodies may be inferred from the
spatio-temporal structures of their magnetic fields. While it is not
yet possible to create convincingly detailed models of planetary
dynamos and of the solar cycle, some major variations of dynamos as function of their
parameters can be explored through numerical simulations. 

The increasing availability in recent years of computer capacity has
facilitated large scale numerical simulations of the generation of
magnetic fields by convection in rotating spherical fluid
shells. Because of limited numerical resolution, molecular values of
material properties are usually not attainable in computer simulations
and eddy diffusivities representing the effects of the unresolved
scales of the turbulent velocity field must therefor be invoked for
comparisons with observations.  It is often assumed for this reason
that the eddy diffusivities for velocities, temperature and magnetic
fields are identical.  The effects of turbulence on the diffusion of
vector and scalar quantities differ, however, and eddy diffusivity
ratios such as the effective Prandtl number and the effective magnetic
Prandtl number thus do not equal unity in general. Besides the Prandtl
numbers, the boundary conditions exert a strong influence on
convection and its dynamo action. Both effects will be considered in
this paper. 

In some respects this paper represents an extension of an earlier
paper (Simitev and Busse, 2005, to which we shall refer to by SB05)
which has focused on the dependence of average properties of
convection driven dynamos on the Prandtl number. New results will be
reported in the following and the emphasis will be placed on time
dependent properties and on dynamo oscillations in particular. In
addition the effect of boundaries of low thermal conductivity will be
studied which are often more realistic than the commonly assumed
boundaries with fixed temperature. Especially in the case of the Earth the 
low conductivity of the mantle will lead to a uniform heat flux from the core unless the 
effect of mantle convection is taken into account. Since the inhomogeneity introduced by the latter
is not well known it will not be considered in the present
analysis. For simulations with various inhomogeneous thermal boundary
conditions see the papers by Glatzmaier \etal~(1999) and by Olson and
Christensen (2002). If compositionally driven convection in the
Earth's core is emphasized the choice of uniform flux at the inner
boundary and of fixed composition at the outer boundary is appropriate
as has been assumed in the work of Glatzmaier and Roberts (1995). 

After a brief introduction of the basic equations and the method of
their numerical solution in section 2, some properties of convection
without magnetic field  will be considered in section 3. In particular
the influence of fixed heat flux boundary conditions will be
explored. In section 4 the onset of convection driven dynamos in
fluids with different Prandtl numbers is described and in section 5
the oscillatory dynamos are interpreted in terms of the theory of
dynamo waves. The influences of various boundary conditions are
studied in section 6 and a concluding discussion is given in the final
section 7.  

\section{Mathematical formulation of the problem and methods of solution}

We consider a rotating spherical fluid shell of thickness $d$
and assume that a  static state exists with the temperature
distribution $T_S = T_0 - \beta d^2 r^2 /2$. 
Here
$\beta=q/(3\,\kappa\,c_p)$ and $T_0=T_1-(T_2-T_1)/(1-\eta)$, where
$T_1$ and $T_2$ are the constant temperatures at the inner and outer
spherical boundaries, $\eta=r_i/r_o$ is the radius ratio of the inner $r_i$ to
the outer $r_o$ radius, $q$ is the uniform heat source density,
$\kappa$ its thermal diffusivity, $c_p$ is its specific heat at constant pressure
and $rd$ is the length
of the position vector with respect to the center of the sphere. 
The gravity field is given by $\vec g = - d \gamma \vec r$. In
addition to  $d$, the time $d^2 / \nu$,  the temperature $\nu^2 /
\gamma \alpha d^4$ and  the magnetic flux density $\nu ( \mu \varrho
)^{1/2} /d$ are used as scales for the dimensionless description of
the problem  where $\nu$ denotes the kinematic viscosity of the fluid,
$\varrho$ its density and $\mu$ is
its magnetic permeability. Since we shall assume the Boussinesq
approximation material properties are regarded as constants except for
the temperature dependence of the density described by $\alpha \equiv
- ( \de \varrho/\de T)/\varrho$ which is taken in the gravity
term. Both, the velocity field $\vec u$ and the magnetic flux 
density $\vec B$, are solenoidal vector fields for which the general
representation 
\begin{subequations}
\begin{align}
&
\vec u = \nabla \times ( \nabla v \times \vec r) + \nabla w \times 
\vec r, \\
&
\vec B = \nabla \times  ( \nabla h \times \vec r) + \nabla g \times 
\vec r,
\end{align}
\end{subequations}
can be employed. By multiplying the (curl)$^2$ and the curl of the
Navier-Stokes equations of motion by $\vec r$ we obtain two equations
for $v$ and $w$,   
\begin{subequations}
\label{momentum}
\begin{align}
&\hspace*{-1.6cm}
[( \nabla^2 - \partial_t) {\cal L}_2 + \tau \partial_{\varphi} ] \nabla^2 v +
\tau {\cal Q} w - {\cal L}_2 \Theta  
= - \vec r \cdot \nabla \times [ \nabla \times ( \vec u \cdot
\nabla \vec u - \vec B \cdot \nabla \vec B)], \\
&\hspace*{-1.6cm}
[( \nabla^2 - \partial_t) {\cal L}_2 + \tau \partial_{\varphi} ] w - \tau {\cal Q}v 
= \vec
r \cdot \nabla \times ( \vec u \cdot \nabla \vec u - \vec B \cdot
\nabla \vec B), 
\end{align}
\end{subequations}
\myfigure{
\mypsfrag{Rt}{\hspace*{-5mm}{${R_c}/{\tau^{4/3}}$}}
\mypsfrag{t}{$\tau$}
\mypsfrag{m}{$m_c$}
\mypsfrag{wt23}{$\omega_c/\tau^{2/3}$}
\mypsfrag{a2}{\hspace*{-2mm}$10^2$}
\mypsfrag{a3}{\hspace*{-2mm}$10^3$}
\mypsfrag{a4}{\hspace*{-2mm}$10^4$}
\mypsfrag{a5}{\hspace*{-2mm}$10^5$}
\mypsfrag{1}{1}
\mypsfrag{10}{10}
\mypsfrag{1.00}{1.0}
\mypsfrag{0.10}{0.1}
\mypsfrag{0.01}{0.01}
\mypsfrag{0.00}{0.001}
\mypsfrag{P02}{}
\mypsfrag{P20}{}
\mypsfrag{P2}{}
\mypsfrag{P02a}{}
\hspace*{-5mm}
\epsfig{file=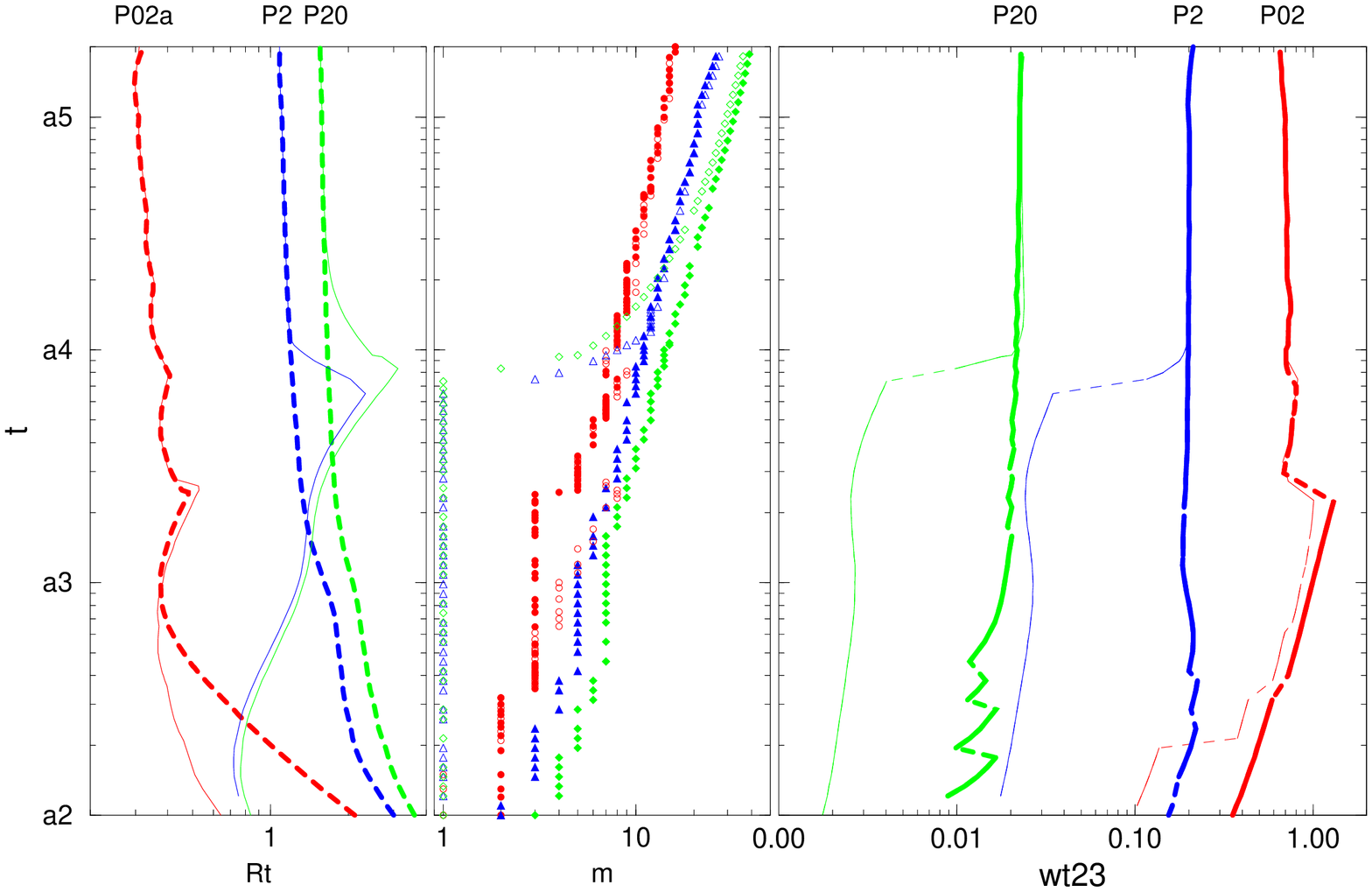,height=13.5cm,angle=-90,clip=}
}{
  Critical values of the Rayleigh number $R_c$,
  wavenumber $m_c$ and frequency $\omega_c$ as functions of $\tau$ in the
  case  $\eta=0.4$ for values of the Prandtl number $P=0.2$ (red), $P=2$
  (blue), $P=20$ (green) and
  UFBC (thick lines, filled symbols) and FTBC (thin lines, empty symbols).  
}{f.0010}

where $\partial_t$ denotes the partial derivative with respect to time
$t$ and where $\partial_{\varphi}$ is the partial derivative with respect to
the angle $\varphi$ of a spherical system of coordinates $r, \theta,
\varphi$. For further details we refer to SB05. The operators ${\cal
L}_2$ and $\cal Q$ are defined by  
\begin{align}
&
{\cal L}_2 \equiv - r^2 \nabla^2 + \partial_r ( r^2 \partial_r), \nonumber\\
&
{\cal Q} \equiv r \cos \theta \nabla^2 - ({\cal L}_2 + r \partial_r ) ( \cos \theta
\partial_r - r^{-1} \sin \theta \partial_{\theta}). \nonumber
\end{align}
The heat equation for the dimensionless deviation $\Theta$ from the
static temperature distribution can be written in the form
\begin{equation}
\label{heat}
\nabla^2 \Theta + R{\cal L}_2 v = P ( \partial_t + \vec u \cdot \nabla ) \Theta,
\end{equation}
and the equations for $h$ and $g$ are obtained through the multiplication of
equation of induction and of its curl by $\vec r$
\begin{subequations}
\label{induction}
\begin{align}
&
\nabla^2 {\cal L}_2 h = P_m [ \partial_t {\cal L}_2 h - \vec r \cdot
\nabla \times ( \vec u \times \vec B )], \\
&
\nabla^2 {\cal L}_2 g = P_m [ \partial_t {\cal L}_2 g - \vec r \cdot
\nabla \times ( \nabla \times ( \vec u \times \vec B ))].
\end{align}
\end{subequations}
The Rayleigh number $R$,
the Coriolis number $\tau$, the Prandtl number $P$ and the magnetic
Prandtl number $P_m$ are defined by 
\begin{equation}
R = \frac{\alpha \gamma \beta d^6}{\nu \kappa} , 
\enspace \tau = \frac{2
\Omega d^2}{\nu} , \enspace P = \frac{\nu}{\kappa} , \enspace P_m = \frac{\nu}{\lambda},
\end{equation}
where $\lambda$ is the magnetic diffusivity. For the static
temperature distribution we have chosen the case of a homogeneously
heated sphere. This state is traditionally used for the analysis of
convection in self-gravitating spheres and offers the numerical
advantage that for Rayleigh numbers close to the critical value $R_c$
the strength of convection does not differ much near the inner and
outer boundaries. As the Rayleigh number increases beyond $R_c$
additional heat enters
at the inner boundary and is delivered by convection to the outer
boundary. When $R$ reaches a high multiple of $R_c$ the heat generated
internally in the fluid becomes negligible in comparison to the heat
transported by convection through the spherical shell.  
\myfigure{
\mypsfrag{t}{$\tau$}
\mypsfrag{0.0}{0.0}
\mypsfrag{0.2}{0.2}
\mypsfrag{0.4}{0.4}
\mypsfrag{0.6}{0.6}
\mypsfrag{0.8}{0.8}
\mypsfrag{1.0}{1.0}
\mypsfrag{0}{0}
\mypsfrag{4}{4}
\mypsfrag{8}{8}
\mypsfrag{4.}{\hspace{-8mm}$4\x10^4$}
\mypsfrag{2}{2}
\mypsfrag{Ex}{$E$}
\mypsfrag{Nu}{$Nu$}
\begin{center}
\epsfig{file=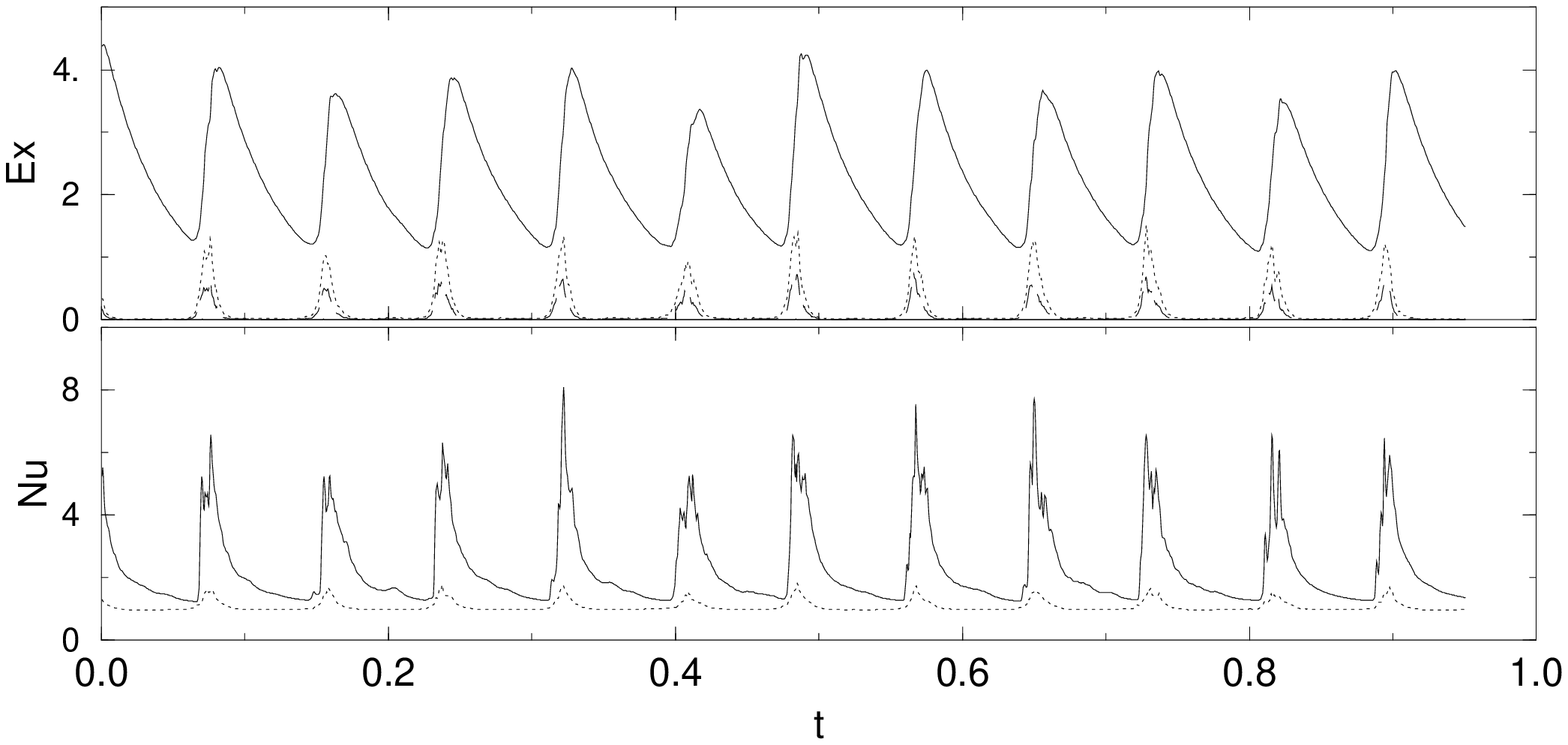,height=13.5cm,width=4.5cm,angle=-90}
\end{center}
}{
Time series of kinetic $E$ energy densities (top) and Nusselt numbers
$Nu$ (bottom) for the convection case $P=0.5$, $\tau=1.5\times10^4$,
$R=1.5\times10^6$ and UFBC.  The components $\overline{E}_t$ and $Nu_o$
are represented by solid, $\check{E}_t$ and $Nu_i$ by dotted lines
and $\check{E}_p$ by a dashed line.}{f.0020}

\myfigure{
\begin{center}
\epsfig{file=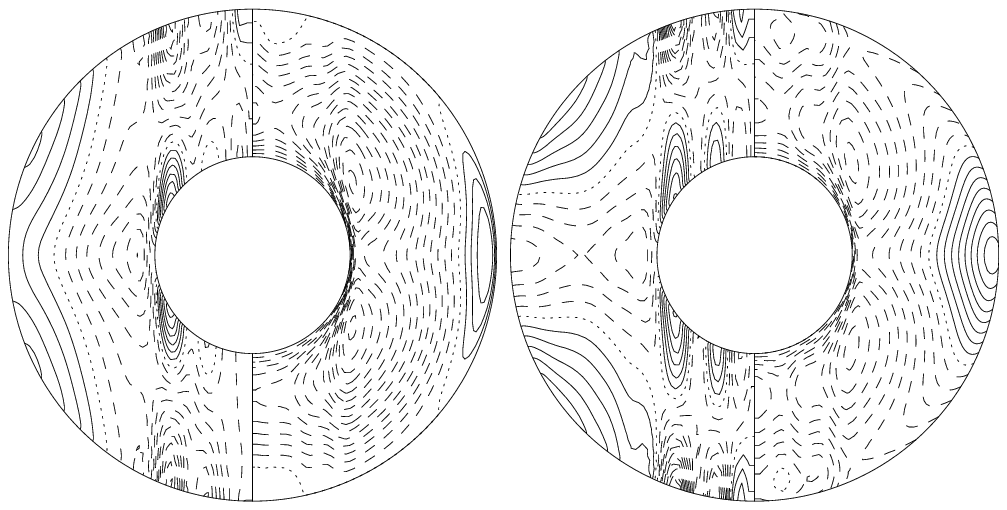,height=3.5cm,clip=}
\end{center}
}{
  Isolines of azimuthally-averaged differential rotation
  $\overline{u_\varphi}$ (left half) and temperature perturbation
  $\overline{\Theta}$ (right half) in the case $P=20$, $\tau=10^4$,
  $R=2\times10^6$ and FTBC (to the left)  or UFBC (to the right). The
  fields represent averages in time.
}{f.0030}

\myfigure{
\mypsfrag{0}{0}
\mypsfrag{30}{30}
\mypsfrag{60}{60}
\mypsfrag{90}{90}
\mypsfrag{0.5}{\hspace*{0mm}0.5}
\mypsfrag{2.5}{\hspace*{0mm}2.5}
\mypsfrag{4.5}{\hspace*{0mm}4.5}
\mypsfrag{1}{1}
\mypsfrag{3}{3}
\mypsfrag{7}{}
\mypsfrag{10}{10}
\mypsfrag{13}{13}
\mypsfrag{-600000}{\hspace{5mm}-6}
\mypsfrag{600000}{}
\mypsfrag{1200000}{\hspace*{5mm}12}
\mypsfrag{No}{$Nu_o$}
\mypsfrag{thhhh}{\hspace*{4mm}$\theta$}
\mypsfrag{a}{(b)}
\mypsfrag{b}{(a)}
\mypsfrag{0}{0}
\mypsfrag{30}{30}
\mypsfrag{60}{60}
\mypsfrag{90}{90}
\mypsfrag{2}{2}
\mypsfrag{4}{}
\mypsfrag{6}{6}
\mypsfrag{8}{8}
\mypsfrag{10}{10}
\mypsfrag{k}{\hspace{-1mm}-4}
\mypsfrag{l}{\hspace{-4mm}-0.5}
\mypsfrag{m}{\hspace*{0mm}3}
\mypsfrag{n}{\hspace*{-2mm}6.5}
\mypsfrag{1}{1}
\mypsfrag{No}{$Nu_o$}
\mypsfrag{thhhh}{\hspace*{4mm}$\theta$}
\mypsfrag{Th}{\hspace*{-7mm}$\overline{\Theta}(r_o) \times 10^{-4}$}
\mypsfrag{bb}{(a)}
\mypsfrag{aa}{(a)}
\mypsfrag{a}{(b)}
\mypsfrag{b}{(b)}
\begin{center}
\hspace*{-4mm}
\epsfig{file=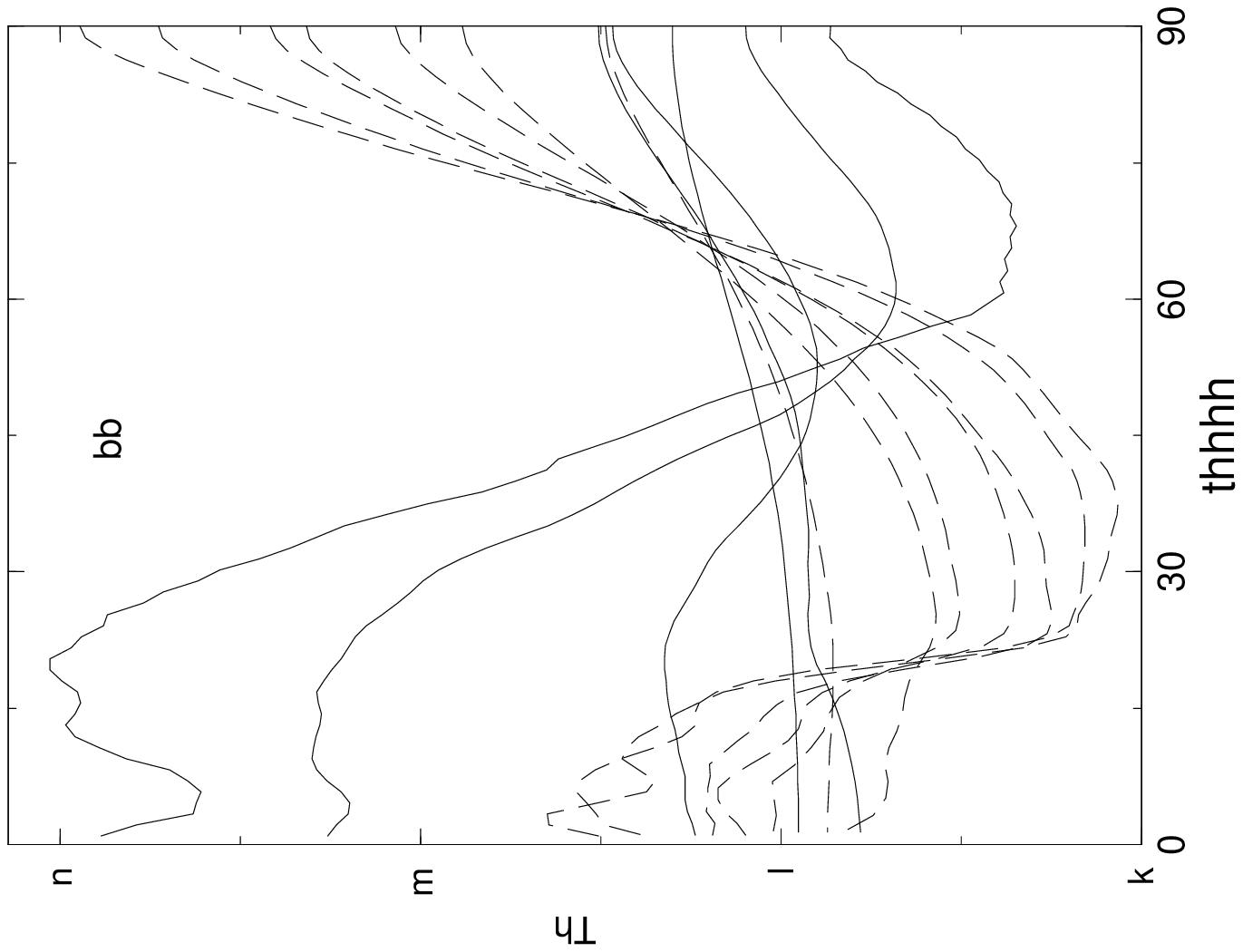,width=8.5cm,angle=-90,clip=}
\epsfig{file=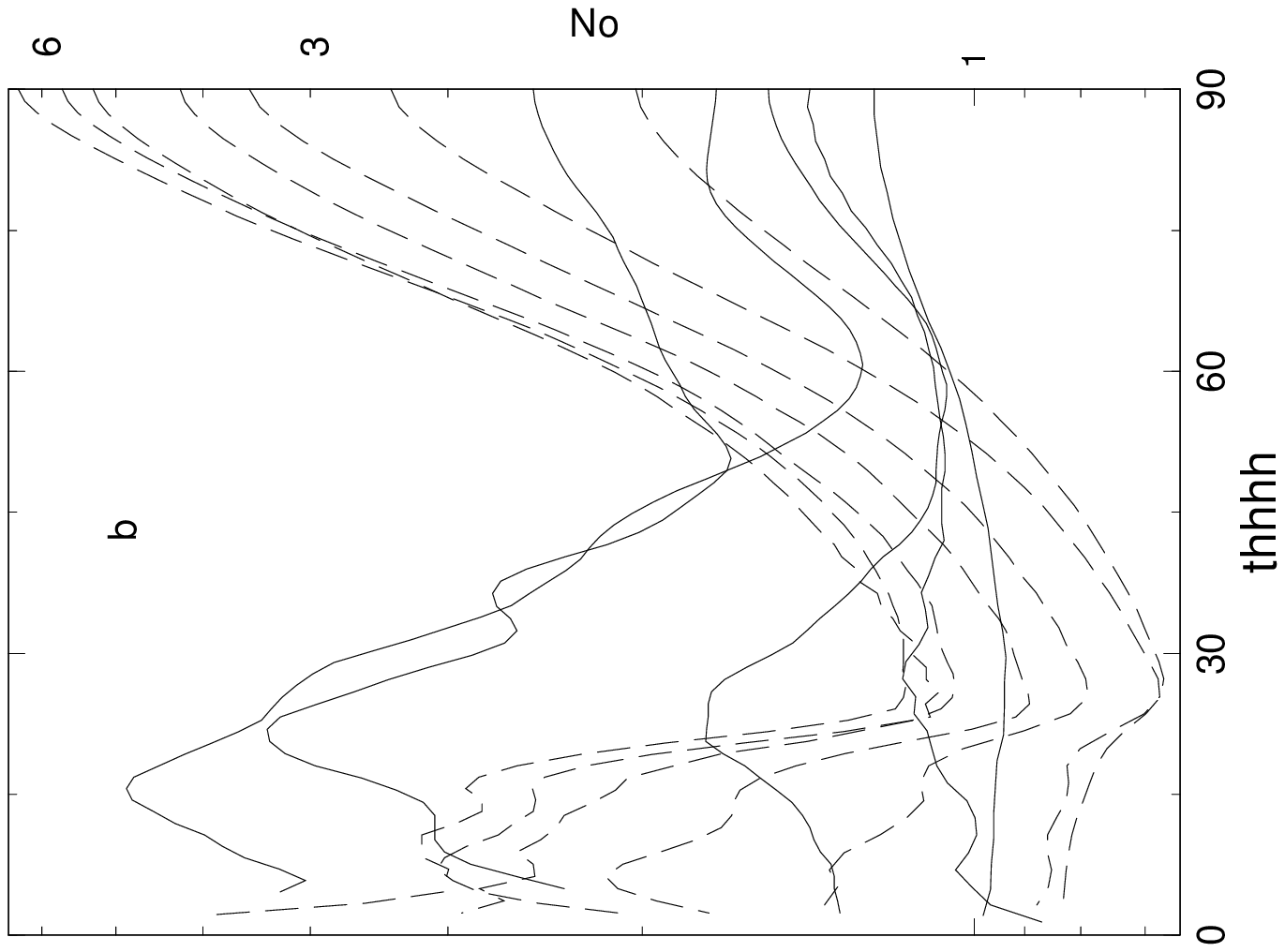,width=8.5cm,angle=-90,clip=}
\end{center}
}{  
Time- and azimuthally-averaged temperature $\overline{\Theta}(r_o, \theta)$ in
the case UFBC (left plot (a)) and Nusselt number $Nu_o(\theta)$ in the
case FTBC (right plot (b)) at the outer boundary as a function of the
colatitude $\theta$ for $P=0.5$ with $\tau=1.5\times10^4$ and 
$R=(10+5n)\times10^5, n=0...4$ from bottom to top at $0^\circ$ (solid lines) and for
$P=20$ with  $\tau  = 10^4$ and $R= (10+5n)\times10^5, n=0...6$ from
bottom to    top at $90^\circ$ (dashed lines).
The temperature values $\overline{\Theta}(r_o,\theta)$ for $P=0.5$
have been divided by 20.
}{f.0050}

\myfigure{
\mypsfrag{a2}{\hspace*{-2mm}$10^2$}
\mypsfrag{a3}{\hspace*{-2mm}$10^3$}
\mypsfrag{a4}{\hspace*{-2mm}$10^4$}
\mypsfrag{a5}{\hspace*{-2mm}$10^5$}
\mypsfrag{Ex}{\hspace*{-2mm}$E$}
\mypsfrag{0.1}{\hspace*{-2mm}0.1}
\mypsfrag{1.0}{\hspace{-2mm}1.0}
\mypsfrag{10.0}{10}
\mypsfrag{10}{10}
\mypsfrag{15}{15}
\mypsfrag{20}{20}
\mypsfrag{25}{25}
\mypsfrag{30}{30}
\mypsfrag{35}{35}
\mypsfrag{40}{40}
\mypsfrag{a}{(a)}
\mypsfrag{b}{(b)}
\mypsfrag{1}{1}
\mypsfrag{10}{10}
\mypsfrag{100}{100}
\mypsfrag{No}{\hspace{-3mm}$Nu$}
\mypsfrag{R}{\hspace{-5mm}$R\times 10^{-5}$}
\begin{center}
\hspace*{4mm}
\epsfig{file=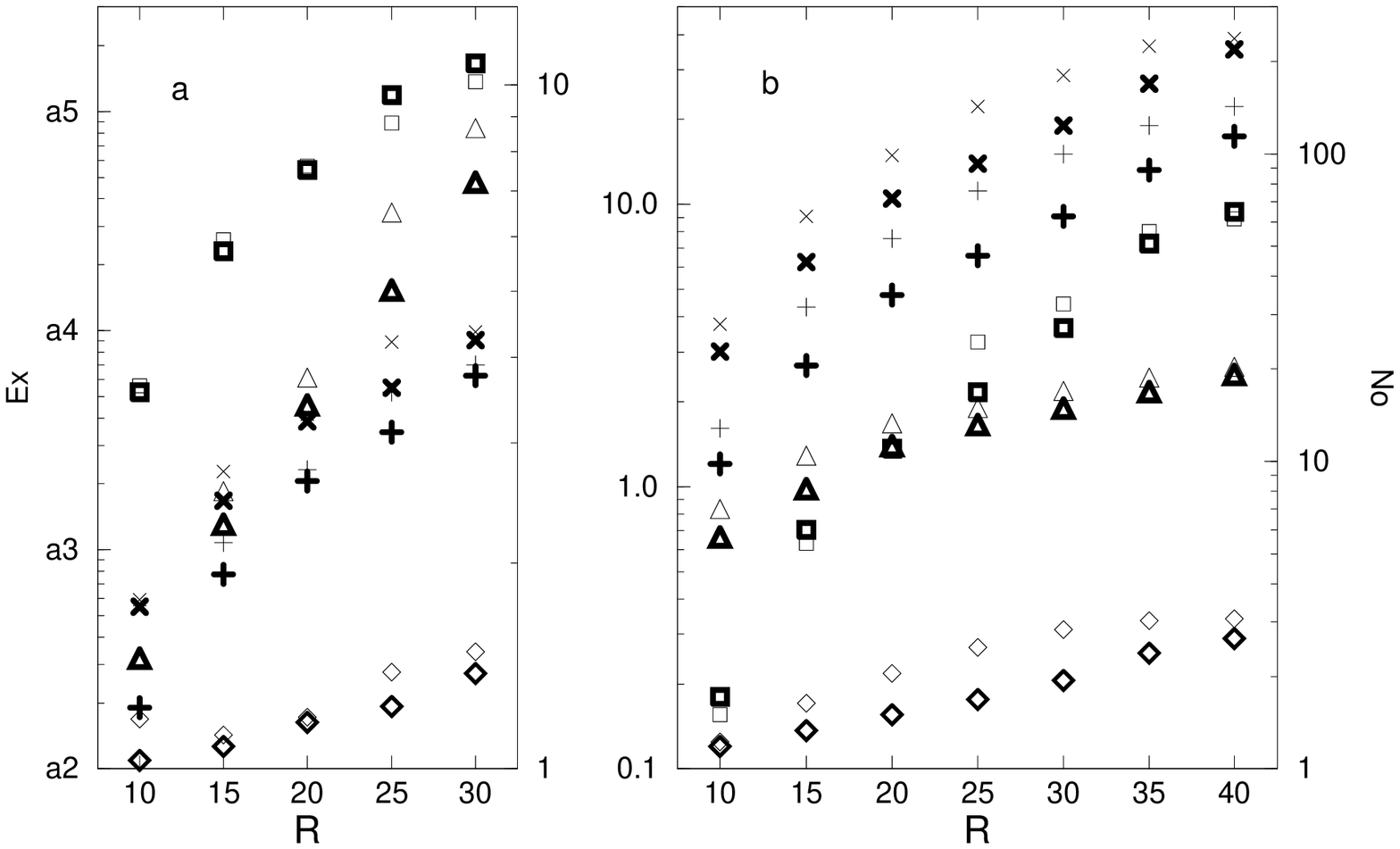,width=8cm,angle=-90}
\end{center}
}{
Kinetic $E$ energy densities (left ordinate)
and $Nu$ (right ordinate) as functions of $R$ for non-magnetic
convection
for (a) $P=0.5$, $\tau=1.5\times10^4$ and
(b) $P=20$, $\tau=10^4$.
The components
$\overline{E}_t$, $\check{E}_p$, $\check{E}_t$ are represented by
squares, plus-signs and crosses, respectively and $Nu_o$ and $Nu_i$ by 
diamonds and triangles. Cases with FTBC are shown with light
symbols, and these with UFBC are shown with heavy symbols.
}{f.0060}

For the velocity field either stress-free boundaries, 
\begin{equation}
\label{vbc}
\hspace*{-8mm}
v = \partial^2_{rr}v = \partial_r (w/r) = 0 
\qquad \mbox{ at } r=r_i \equiv \eta/(1-\eta) \mbox{ and } r=r_o \equiv 1/(1-\eta),
\end{equation}
or no-slip boundaries,
\begin{equation}
\label{vbd}
\hspace*{-8mm}
v = \partial_r v = w = 0 
\qquad \mbox{ at } r=r_i \equiv \eta/(1-\eta) \mbox{ and } r=r_o \equiv 1/(1-\eta),
\end{equation}
will be assumed. 
Unless indicated otherwise $\eta=0.4$  and stress-free
boundary conditions will be used in
the following. 
The value $\eta=0.4$ is not only employed because it turns out to be close to the
radius ratio of the liquid outer core of the Earth, but because it
also provides a good balance between the regions inside and outside the
tangent cylinder touching the inner boundary at its equator. These two
regions are known, of course, for their rather different dynamical properties. 

For the thermal boundary conditions we shall assume
that either the temperature is fixed (FTBC, i.e.~Fixed Temperature
Boundary Conditions), 
\begin{equation}
\label{vbe}
\hspace*{-8mm}
\Theta = 0 
\end{equation}
or its normal derivative is fixed (UFBC, i.e.~Uniform Flux Boundary Conditions),
\begin{equation}
\label{vbf}
\hspace*{-8mm}
\frac{\partial}{\partial r}(\Theta - \overline{\overline{\Theta}}) = 0 \mbox{ , but}\quad  \overline{\overline{\Theta}} = 0 
\end{equation}
at the boundary where the double overbar indicates the average over
the spherical boundary. This latter definition has been used in order
that the Rayleigh number continues to be based on the difference
between the average temperatures between the boundaries. 

For the magnetic field electrically insulating
boundaries are assumed such that the poloidal function $h$ must be 
matched to the function $h^{(e)}$ which describes the  
potential fields 
outside the fluid shell  
\begin{equation}
\hspace*{-8mm}
\label{mbc}
g = h-h^{(e)} = \partial_r ( h-h^{(e)})=0 
\qquad \mbox{ at }r \equiv \eta/(1-\eta) \mbox{ and } r \equiv 1/(1-\eta).
\end{equation}
But computations for the case of an inner boundary with no-slip
conditions and an electrical conductivity equal to that of the fluid
have also been done. The numerical integration of equations
\eqref{momentum},\eqref{heat} and \eqref{induction} together with boundary
conditions \eqref{vbc} or \eqref{vbd} and \eqref{vbe} or \eqref{vbf}
and \eqref{mbc} proceeds with the pseudo-spectral  
method as described by Tilgner and Busse (1997) and Tilgner (1999)
which is based on an expansion of all dependent variables in
spherical harmonics for the $\theta , \varphi$-dependences, i.e. 
\begin{equation}
v = \sum \limits_{l,m} V_l^m (r,t) P_l^m ( \cos \theta ) \exp \{ im \varphi \}
\end{equation}
and analogous expressions for the other variables, $w, \Theta, h$ and $g$. 
$P_l^m$ denotes the associated Legendre functions.
For the $r$-dependence expansions in Chebychev polynomials are used. 
For further details see also Busse {\it et al.} (1998) or Grote {\it  et al.} (2000).
For the computations to be reported in the following a minimum of
33 collocation points in
the radial direction and spherical harmonics up to the order 64 have been
used. But in many cases the resolution has been increased to 49 collocation
points and spherical harmonics up to the order 96 or 128.

\section{Convection in rotating spherical shells}

For an introduction to the problem of convection in spherical shells 
we refer to the recent review of Busse (2002a). Additional information
can be found in the papers by Grote and Busse (2001), Christensen
(2002) and Simitev and Busse (2003, 2005). Here we shall focus the
attention on the case of uniform heat flux boundary conditions. In figure \ref{f.0010}
the critical value $R_c$ of the Rayleigh number and the associated
critical azimuthal wavenumber $m_c$ have been plotted as a function of
$\tau$ for several values of the Prandtl number $P$. The results for
the case of fixed temperatures are shown for comparison. In the case
of the steady onset of convection in a planar layer heated from below
the critical value of the Rayleigh number and the associated
wavenumber decrease monotonously with a decreasing thermal
conductivity of a boundary (Sparrow \etal , 1964). There the onset of
convection is described by a one-dimensional eigenvalue problem, while
in the present case a two-dimensional problem in the complex domain
must be solved because of the time dependent onset. Hence exceptions
from the monotonous decrease of the critical values $R_c, m_c$ in
going from the fixed temperature to fixed flux boundary condition must
be expected as are indeed found in
the intermediate range of $\tau$, $10^3<\tau <3\times 10^4$. This
phenomenon can also be recognized in the analytical results of Busse
and Simitev (2004) obtained in the low Prandtl number limit. As the
critical wavenumber $m_c$ increases the value $R_c$ becomes
independent of the boundary conditions as must be anticipated since the
$r$-dependence becomes negligible in comparison with the azimuthal
dependence of convection. The same independence is, of course, also
evident in the asymptotic theory of Busse (1970).

At finite amplitudes convection in the presence of uniform heat flux
boundaries evolves in a similar way as convection with fixed
temperatures at the boundaries as long as the wavenumbers $m_c$ of the
two cases do not differ significantly. It should be mentioned here
that in contrast to the linear results displayed in figure \ref{f.0010} only an
outer uniform-flux boundary has been assumed for the computations of
finite amplitude convection and its dynamos, while the fixed
temperature condition has been kept at the inner boundary. This
combination seems to be most appropriate for some planetary
applications as mentioned in the Introduction.  
\myfigure{
\mypsfrag{aa01}{$10^{-1}$}
\mypsfrag{aa0}{$10^0$}
\mypsfrag{aa1}{$10^1$}
\mypsfrag{Ex}{$E_x$, $M_x$}
\mypsfrag{1aa0}{\hspace{0mm}$10^0$}
\mypsfrag{1aa1}{\hspace{0mm}$10^1$}
\mypsfrag{1aa2}{\hspace{0mm}$10^2$}
\mypsfrag{1aa3}{\hspace{0mm}$10^3$}
\mypsfrag{2aa1}{$10^1$}
\mypsfrag{2aa2}{$10^2$}
\mypsfrag{2aa3}{$10^3$}
\mypsfrag{2aa4}{}
\mypsfrag{2aa5}{$10^5$}
\mypsfrag{2aa6}{$10^6$}
\mypsfrag{5}{5}
\mypsfrag{10} {10}
\mypsfrag{3}  {3}
\mypsfrag{1} {1}
\mypsfrag{12}  {12}
\mypsfrag{15}  {15}
\mypsfrag{20}  {20}
\mypsfrag{30}  {30}
\mypsfrag{30}  {30}
\mypsfrag{0} {0}
\mypsfrag{c2}{\hspace{-2mm}$5$}
\mypsfrag{f3}{\hspace{-7mm}$10$}
\mypsfrag{f5}{\hspace{5mm}$20$}
\mypsfrag{d} {\hspace{-2mm}$5$}
\mypsfrag{g3}{\hspace{-7mm}$10$}
\mypsfrag{g5}{\hspace{5mm}$20$}
\mypsfrag{Vxxx}{\hspace{-1mm}$V_x$, $O_x$}
\mypsfrag{R}{}
\mypsfrag{RR}{\hspace{10mm}$P$}
\begin{center}
\hspace*{-4mm}
\begin{tabular}{@{}c@{\extracolsep{3mm}}c}
\epsfig{file=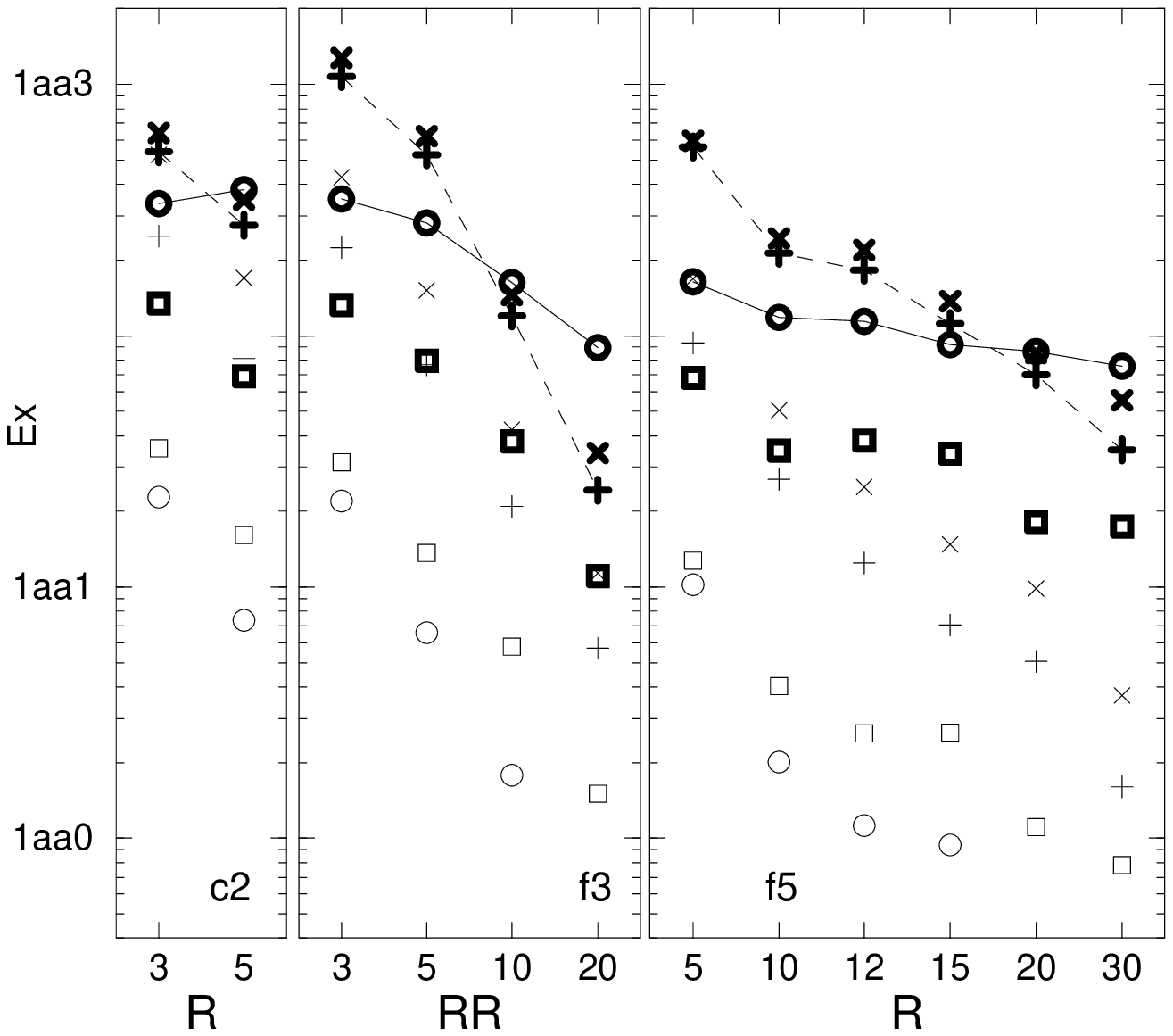,width=8cm,height=7cm,angle=-90} &
\epsfig{file=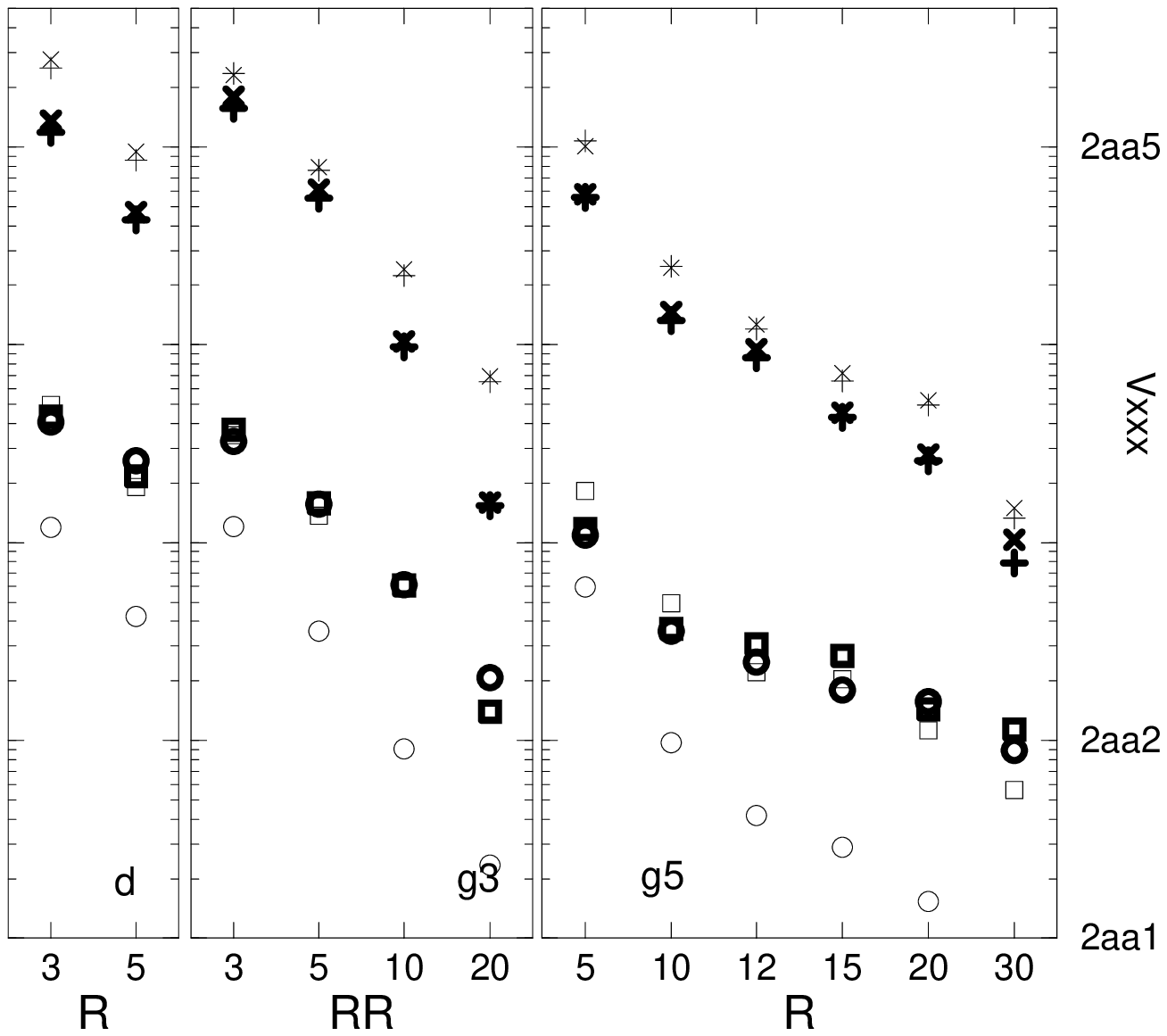,width=8cm,height=7cm,angle=-90} \\
\end{tabular}
\end{center}
}{
Kinetic $E_x$ and magnetic $M_x$ energy densities (left)
and viscous $V_x$ and Ohmic $O_x$ dissipations (right) as
functions of $P$ for convection driven dynamos
for $\tau=10^4$, $R=2\times10^6$, FTBC and magnetic Prandtl  number
as indicated in the boxes.
The components $\overline{X}_p$, $\overline{X}_t$,
$\check{X}_p$, $\check{X}_t$ (where $X = E$, $M$, $V$, $O$) are
represented by circles, squares, plus-signs and
crosses, respectively. Kinetic energy densities and viscous
dissipations are shown with light symbols, magnetic energy
densities and Ohmic dissipations are shown with heavy
symbols. $\overline{E}_p$ is multiplied by a factor 20.
}{f.0080}

\myfigure{
\mypsfrag{o}{\hspace*{-.8mm}$\odot$}
\psfrag{Q}{\hspace*{-0.55mm}$\blacktriangle$}
\psfrag{D}{\hspace*{-0.55mm}$\triangledown$}
\mypsfrag{H}{\hspace*{-.8mm}$\triangle$}
\mypsfrag{M}{\hspace*{-.4mm}$\blacksquare$}
\mypsfrag{P}{$P$}
\mypsfrag{Pm}{$P_m$}
\mypsfrag{R}{$R\times10^{-5}$}
\mypsfrag{b0}{0} 
\mypsfrag{b5}{5} 
\mypsfrag{b10}{10}
\mypsfrag{b15}{15}
\mypsfrag{b20}{20}
\mypsfrag{1}{1}  
\mypsfrag{3}{3}  
\mypsfrag{5}{5}  
\mypsfrag{10}{10} 
\mypsfrag{20}{20}
\mypsfrag{50}{50}
\npsfrag{a1}{tl}{1}  
\npsfrag{a3}{tl}{3}  
\npsfrag{a5}{tl}{5}  
\npsfrag{a10}{tl}{10}
\npsfrag{a20}{tl}{20}
\npsfrag{a50}{tl}{50} 
\npsfrag{a100}{tl}{100}
\npsfrag{a200}{tl}{200}
\begin{center}
\hspace*{10mm}
\epsfig{file=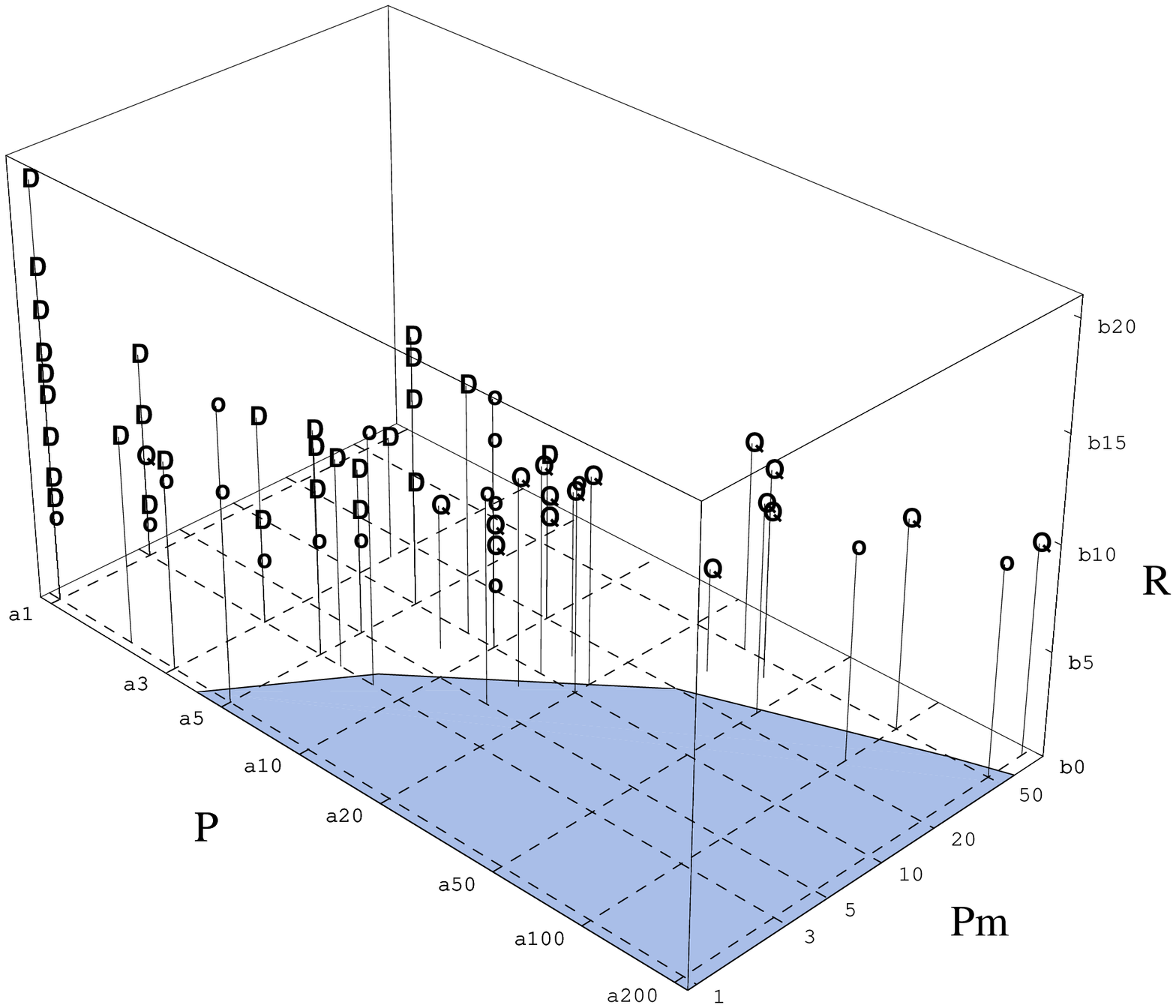,width=12.5cm}
\end{center}
}{
Convection-driven dynamos as function of $R$, $P$ and $P_m$ for
$\tau=5\times10^3$. The symbols indicate dynamos for which
$\overline{M}_p < \check{M}_p$ ($\triangledown$), those for which
opposite is true ($\blacktriangle$) and decaying dynamos ($\odot$).
}{f.0090}

The initial regime of drifting thermal Rossby waves is followed by the
subsequent stages of vacillating convection, localized convection and
relaxation oscillation that are observed with increasing Rayleigh
number for Prandtl numbers of the order unity or less (but above
values of $P$ where convection in the form of inertial waves
dominates, i.e. $P > 10/\sqrt{\tau}$ according to Ardes et al. (1996)). No significant influence of the thermal boundary
conditions can be noticed. The relaxation oscillations shown in figure~\ref{f.0020}
seem to be a little more pronounced than in the case of a fixed
temperature outer boundary, but are otherwise identical. At higher
Prandtl numbers a stronger tendency towards the thermal wind type of
differential rotation can be noticed in the presence of a uniform-flux
thermal boundary as seen in figure \ref{f.0030}. This is a consequence of the
stronger latitudinal temperature gradient admitted by the insulating
boundary condition. 

At finite amplitudes of convection the heat transport is of special
interest the efficiency of which is measured by the Nusselt
number. This is defined as the heat transport in the presence of
convection divided by the heat transport in the absence of motion. In
the case of the spherical fluid shell two Nusselt numbers can be
defined measuring the efficiency of convection at the inner and the
outer boundary, 
\begin{equation}
Nu_i=1- \frac{P}{r_i} \left.\frac{d \overline{\overline{\Theta}}}{d r}\right|_{r=r_i}\qquad Nu_o=1- \frac{P}{r_o} \left.\frac{d \overline{\overline{\Theta}}}{d r}\right|_{r=r_o}
\end{equation}  
where the double bar indicates the average over the spherical surface. In addition the local Nusselt numbers as function of latitude
\begin{equation}
Nu_i(\theta) =1- \frac{P}{r_i} \left.\frac{d\overline{\Theta}}{d r}\right|_{r=r_i}\qquad Nu_o(\theta) =1- \frac{P}{r_o} \left.\frac{d\overline{\Theta}}{d r}\right|_{r=r_o}
\end{equation}  
are of interest where only the azimuthal average is applied, as
indicated by the single bar. The latitudinal dependence of the local
Nusselt number and of the azimuthally averaged temperature at the outer
boundary are shown in figure \ref{f.0050} for the cases of fixed temperature
conditions (right) and uniform flux conditions (left) in the cases $P=0.5$ and $P=20$. As must be
expected latitudes of high temperatures at the boundary correspond to
those with high values of  
$Nu_o(\theta)$.  This figure demonstrates that at low supercritical
Rayleigh numbers the heat transport occurs primarily across the
equatorial region, but as $R$ increases the heat transport in the
polar regions takes off, especially in the case of the lower value of
P, and soon exceeds that at low latitudes. In the polar 
regions convection is better adjusted for carrying heat from the lower
boundary to the upper one, and it is known from computations of the
convective heat transport in horizontal layers rotating about a
vertical axis that the value of $Nu$ may exceed the value in a
non-rotating layer at a given value of $R$ in spite of the higher
critical Rayleigh number in the former case. 

At higher values of the Prandtl number the takeover of the heat
transport by convection in the polar regions is less pronounced, but
the contrast between the equatorial region and 
the mid-latitudes is even stronger than for lower values of
$P$. Again, a dramatic difference between fixed flux and fixed
temperature boundary conditions can not be discerned.  
\myfigure{
\mypsfrag{10}{\hspace{-1mm}10}
\mypsfrag{P}{$P$}
\mypsfrag{Pm}{$P_m$}
\begin{center}
\hspace*{-8mm}
\epsfig{file=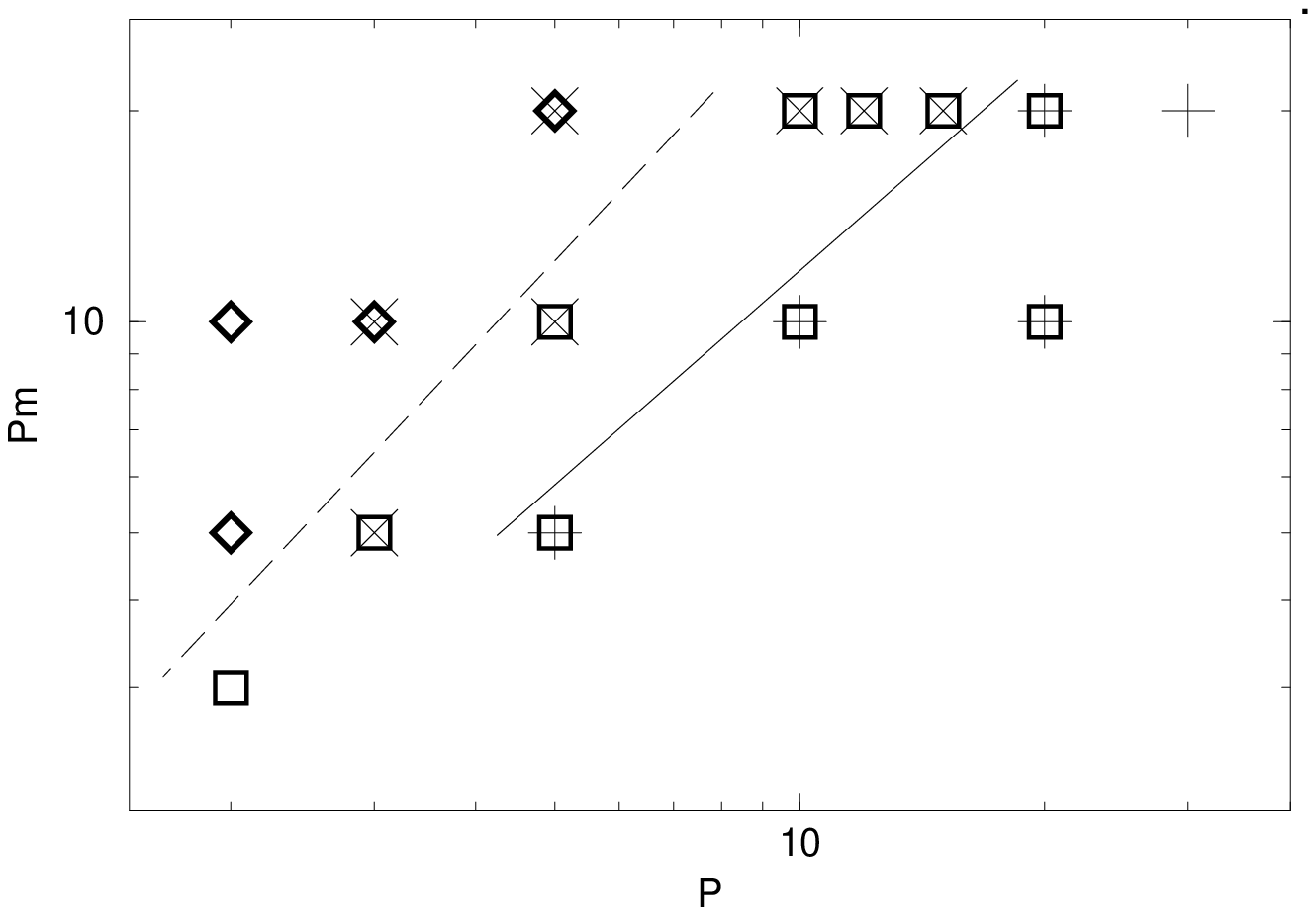,height=9cm,angle=-90}
\end{center}
}{
Convection-driven dynamos as function of $P$ and $P_m$ with $\tau=10^4$
and $R=2\times10^6$ (light symbols) and with $\tau=3\times 10^4$ and
$R=3.5\times10^6$ (heavy symbols). The dynamos for which
$\overline{M}_p>\check{M}_p$ are indicated by squares in the case of
$\tau=3\times 10^4$ and by plus-signs in the case of $\tau=10^4$. The
dynamos for which $\overline{M}_p < \check{M}_p$ are indicated by
diamonds in the case of $\tau=3\times 10^4$ and by crosses in the case
of $\tau=10^4$. The broken and the solid lines represent the
approximate location of the parameter values where the transition from
$\check{M}_p$-dominated to $\overline{M}_p$-dominated dynamos occurs for
$\tau=3\times 10^4$ and $\tau=10^4$, respectively.
}{f.0090.1}

It is also useful to compare the averages over space and time of the
kinetic energy densities of the various components of the convection
flow. They are defined by
\begin{subequations}
\label{edens}
\begin{align}
&
\overline{E}_p = \frac{1}{2} \langle \mid \nabla \times ( \nabla \bar v \times \vec r )
\mid^2 \rangle , \quad \overline{E}_t = \frac{1}{2} \langle \mid \nabla \bar w \times
\vec r \mid^2 \rangle, \\
&
\check{E}_p = \frac{1}{2} \langle \mid \nabla \times ( \nabla \check v \times \vec r )
\mid^2 \rangle , \quad \check{E}_t = \frac{1}{2} \langle \mid \nabla \check w \times
\vec r \mid^2 \rangle,
\end{align}
\end{subequations}
where the angular brackets indicate the average over the fluid shell and over time
and $\bar v$ refers to the azimuthally averaged component of $v$,
while $\check v$ is defined by $\check v = v - \bar v $. In figure \ref{f.0060}
the growth of these energies with the the Rayleigh number is shown for
both, the case with an insulating and the case with a fixed
temperature at the outer boundary.  
In the low Prandtl number case of the left diagram the rapid growth with $R$ of $\overline{E}_t$ corresponding to the energy of
differential rotation is remarkable. Only for higher values of $P$ does $\overline{E}_t$ 
never exceed the energies of the fluctuating components of
motion.
\myfigure{
\mypsfrag{E}{$E$}
\mypsfrag{R}{$R$}
\mypsfrag{a6}{$10^6$}
\mypsfrag{a-4}{\hspace*{-1mm}$10^{-4}$}
\mypsfrag{a-3}{}
\mypsfrag{a-2}{\hspace*{-1mm}$10^{-2}$}
\mypsfrag{a-1}{}
\mypsfrag{a0}{\hspace*{-1mm}$10^0$}
\mypsfrag{a1}{}
\mypsfrag{aa}{\hspace{-3mm}$4\times10^5$}
\mypsfrag{10}{10}
\mypsfrag{1}{1}
\mypsfrag{Ni}{$Nu_i$}
\mypsfrag{40}{40}
\mypsfrag{20}{20}
\mypsfrag{10}{10}
\mypsfrag{15}{15}
\mypsfrag{Pm}{$P_m$}
\begin{center}
\hspace*{-3mm}
\epsfig{file=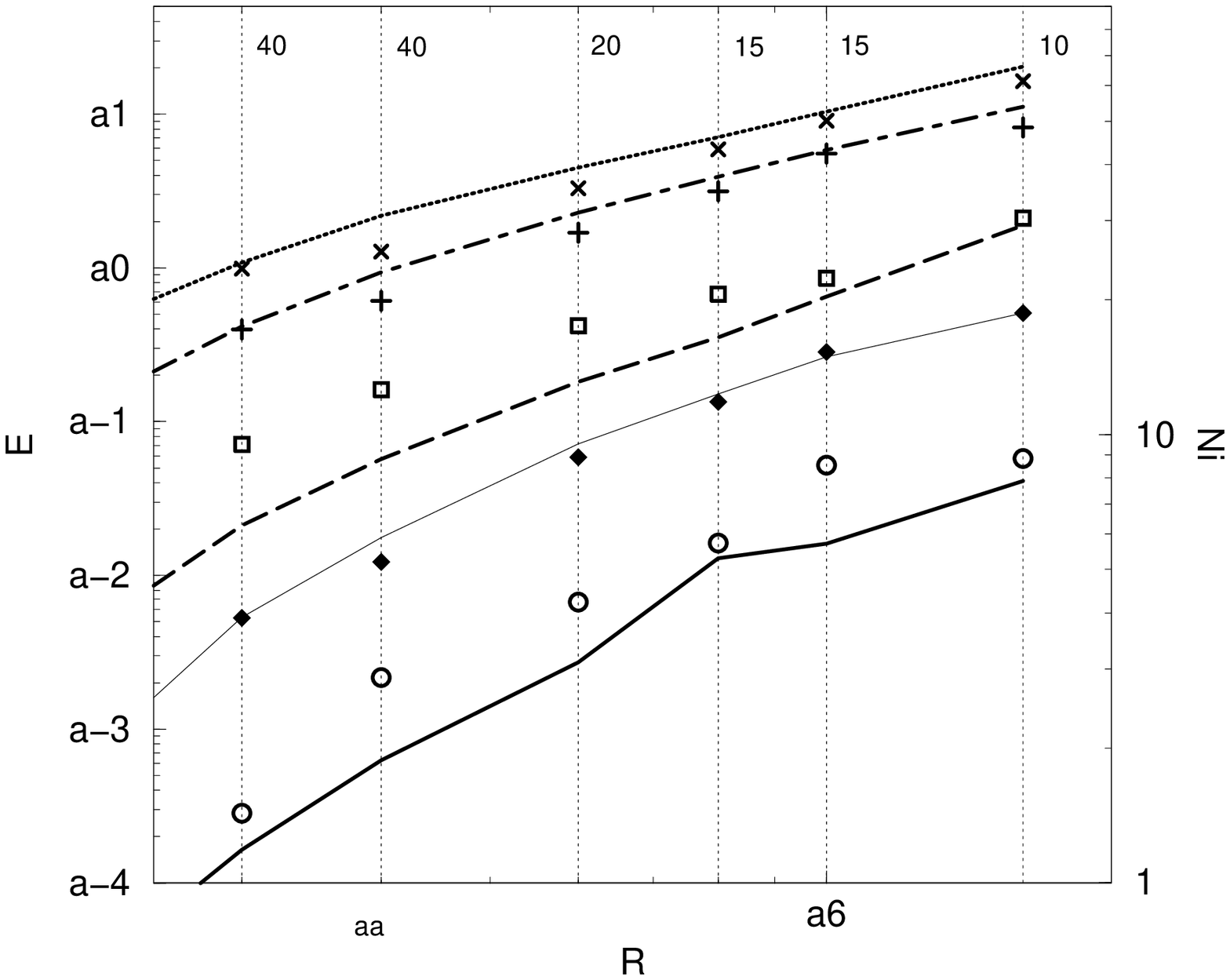,height=10cm,angle=-90,clip=}
\end{center}
}{
Time-averaged kinetic energy densities $\overline{E}_p$, (solid line
and circles), $\overline{E}_t$ (dashed line and squares),
$\check{E}_p$ (dashed-dotted line and plus-signs), $\check{E}_t$
(dotted line and crosses)  and Nusselt number
$Nu_i$ (thin line and solid diamonds) of non-magnetic convection
(lines) and dynamo solutions (symbols) for $P=20$,
$\tau=5\times10^3$ and $R$ as indicated at the x-axis. The symbols
connected by the dotted lines correspond to the values of $P_m$ given
at the top of the figure. 
}{f.0100}

\myfigure{
\begin{center}
\epsfig{file=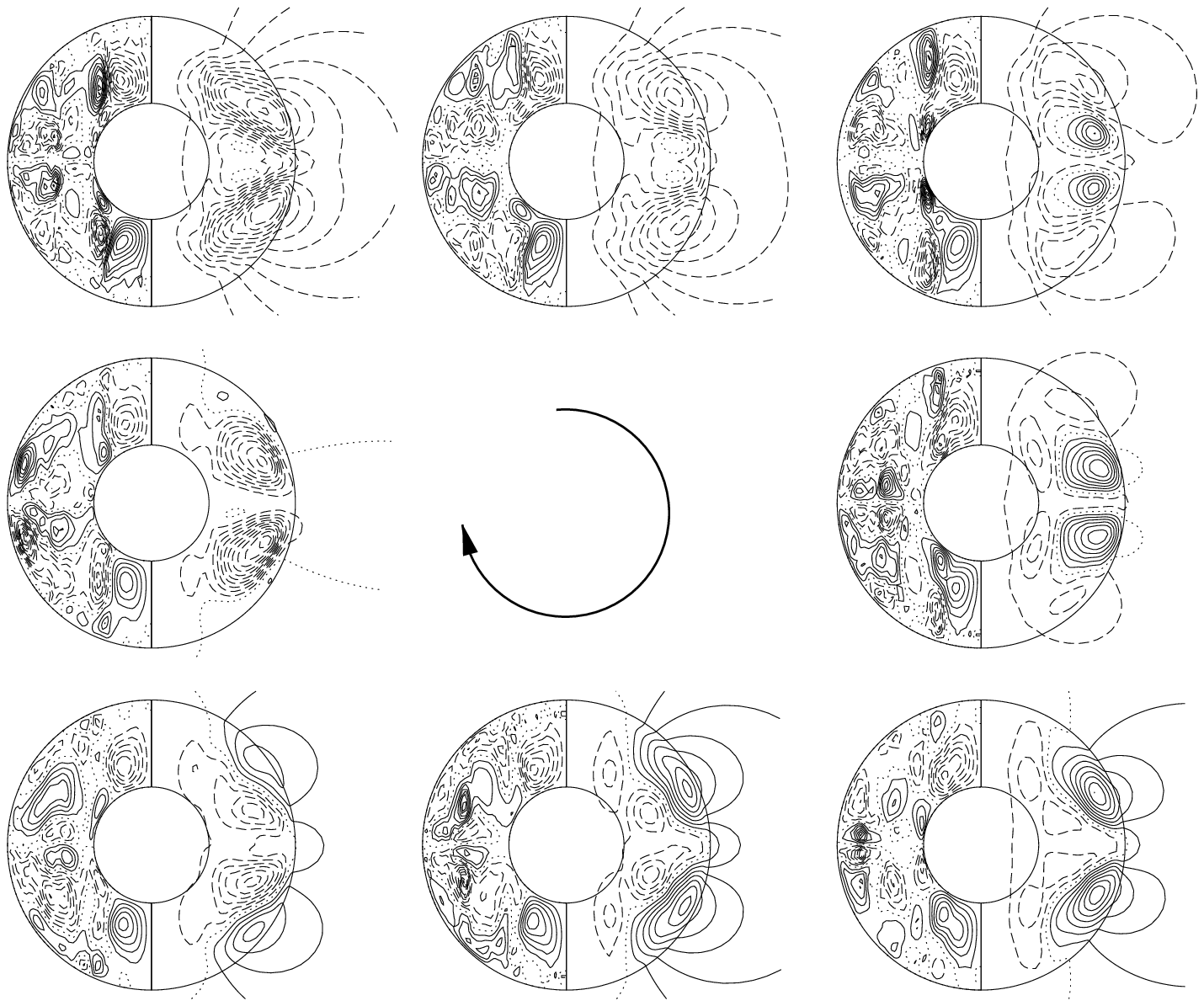,width=11cm,angle=0,clip=}
\end{center}
}{
A period of dipolar oscillations for $P=5$, $\tau=5\times10^3$,
$R=6\times10^5$, $Pm=10$ with $\Lambda =0.1429$. The   plots start at the upper left corner
and follow clockwise with $\Delta t
=0.035$ and show  meridional lines of constant
$\overline{B_{\varphi}}$ in the left half and of $r \sin \theta
\dd_\theta \overline{h}=$const.~in the right half of each circle.
}{f.0130}

\myfigure{
\begin{center}
\mypsfrag{60}{60}
\mypsfrag{65}{65}
\mypsfrag{70}{70}
\mypsfrag{4.8}{4.8}
\mypsfrag{5}{5}
\mypsfrag{5.4}{5.4}
\mypsfrag{5.5}{5.5}
\mypsfrag{5.6}{5.6}
\mypsfrag{6}{6}
\mypsfrag{8}{8}
\mypsfrag{9}{9}
\mypsfrag{11}{11}
\mypsfrag{14}{14}
\mypsfrag{15}{15}
\mypsfrag{17}{17}
\mypsfrag{20}{20}
\mypsfrag{22}{22}
\mypsfrag{23}{23}
\mypsfrag{26}{26}
\mypsfrag{27}{27}
\mypsfrag{5}{5}
\mypsfrag{6}{6}
\mypsfrag{8}{8}
\mypsfrag{.11}{\hspace{-1mm}.11}  
\mypsfrag{.12}{\hspace{1mm}.12} 
\mypsfrag{.1}{.1}  
\mypsfrag{.15}{.15}  
\mypsfrag{2}{2} 
\mypsfrag{1}{1}  
\mypsfrag{0.8}{0.8}  
\mypsfrag{3}{3} 
\mypsfrag{5}{5}  
\mypsfrag{4}{4}  
\mypsfrag{6}{6} 
\mypsfrag{10}{10}  
\mypsfrag{1}{1}  
\mypsfrag{0.6}{0.6} 
\mypsfrag{0.5}{\hspace{-1mm}.5}
\mypsfrag{0.4}{0.4}
\mypsfrag{0.2}{0.2} 
\mypsfrag{0.0}{0.0}
\mypsfrag{0.00}{0.00}
\mypsfrag{0.02}{0.02}
\mypsfrag{0.04}{0.04}
\mypsfrag{0.06}{0.06}
\mypsfrag{T}{\hspace{-30mm}$T$} 
\mypsfrag{Tb}{\hspace{30mm}$T$} 
\mypsfrag{Ta}{}  
\mypsfrag{Pm}{\hspace{-7mm}$P_m$} 
\mypsfrag{R}{\hspace{-12mm}$R\times10^{-5}$}  
\mypsfrag{a}{\hspace{-2mm}(a)}  
\mypsfrag{b}{(b)}  
\mypsfrag{c}{\hspace{-1mm}(c)} 
\mypsfrag{d}{\hspace{-2mm}(d)}  
\mypsfrag{e}{\hspace{-2mm}(e)} 
\mypsfrag{f}{\hspace{-4mm}(f)}  
\hspace*{-3mm}
\hspace*{-5mm}
\epsfig{file=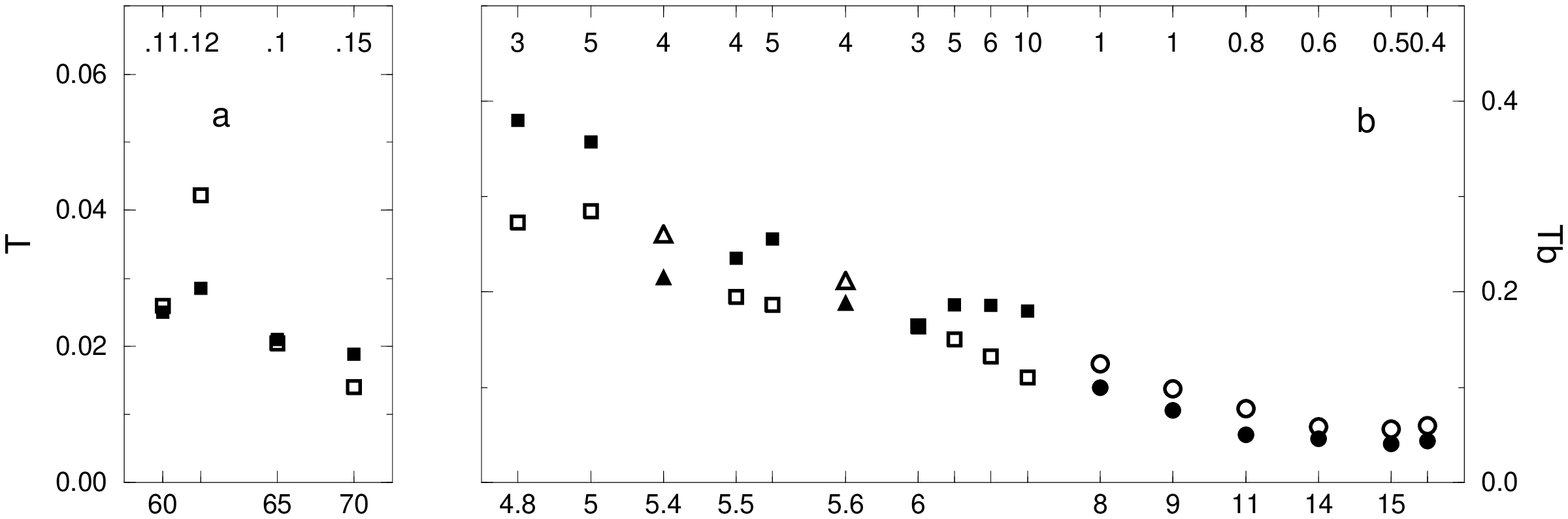,height=14.5cm,angle=-90}\\[5mm]
\hspace*{-6.9mm}
\epsfig{file=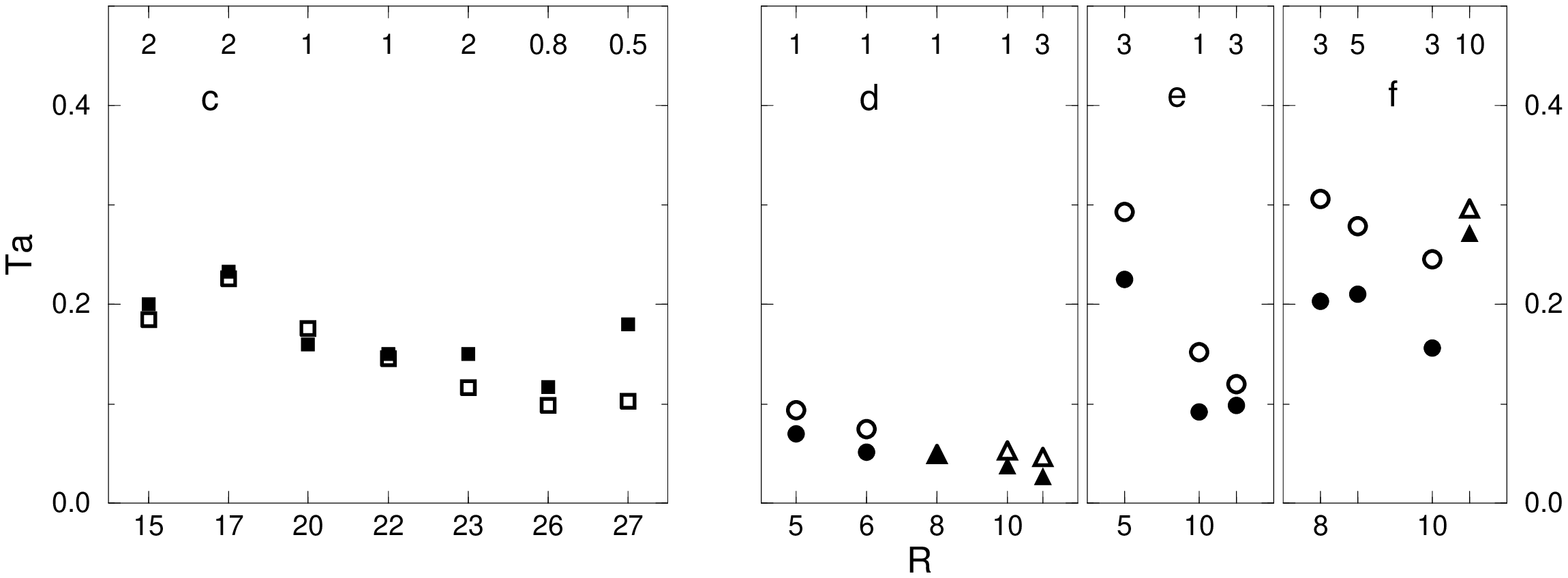,height=13.9cm,angle=-90}
\end{center}
}{
Period of dynamo oscillations $T$ obtained from the analytical expression
(5.8) (empty symbols) and from simulations (solid symbols)
for (a) $P=0.1$, $\tau=10^5$, (b) $P=1$,
$\tau=10^4$, (c) $P=1$, $\tau=3\times10^4$ and (d) 
$P=1$, (e) $P=3$, (f) $P=5$ and $\tau=5\times10^3$. The values of $R$ 
are given at the $x$-axis and the values of $P_m$ are given at the
top of the panels.
The symbols indicate quadrupolar (circles),
hemispherical (squares) and dipolar (triangles) dynamos.
}{f.0120}

\myfigure{
\begin{center}
\epsfig{file=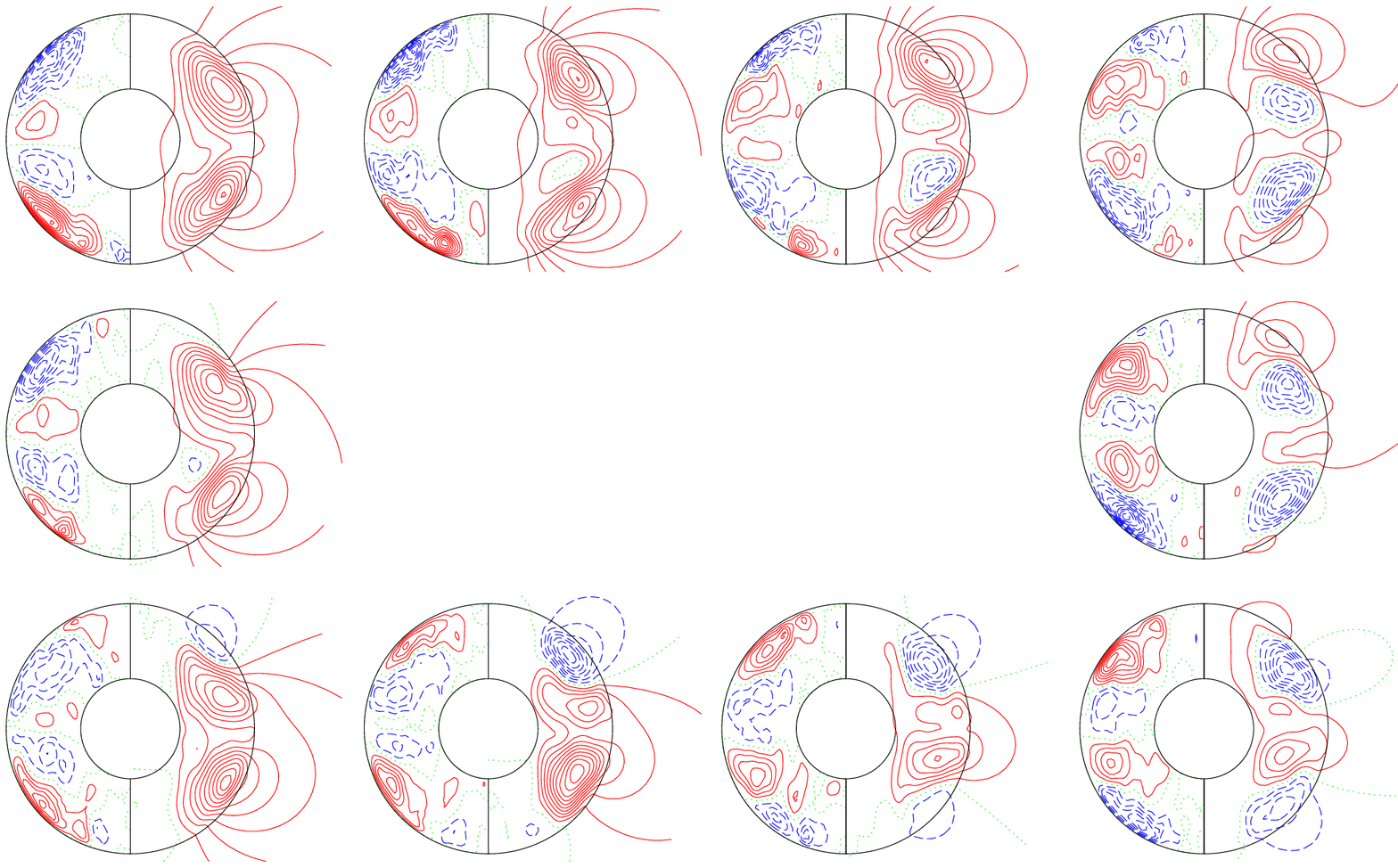,width=14cm,clip=}
\end{center}
}{
A period of dipolar oscillations for
$P=1$,  $\tau=5\times10^3$, $R=10^6$, $Pm=1$ with $\Lambda=0.6794$. The plots start at the upper
left corner and follow clockwise with $\Delta t
  =0.0051$ and show isolines similar to those in figure \ref{f.0130}.
}{f.0140}

\section{Convection-driven dynamos}

The onset convection-driven dynamos as function of parameters
equivalent to those given by expressions (2.5) has been investigated
in numerous papers, see, for example, Busse \etal~(1998), Christensen
\etal~(1999), Grote \etal~(2000, 2001), Grote and  
Busse (2001), Kutzner and Christensen (2000, 2002), SB05. For displays of dynamo onsets as functions of $P, R$ and $\tau$ see SB05 and Busse (2002). Typically the Rayleigh number required for dynamos increases with
decreasing $P_m$ such that the minimum magnetic Reynolds number $Rm$
defined by  $Rm = (2E)^{1/2}P_m$ stays constant at a 
value of the order $10^2$.  As has been found before (Christensen
\etal \;1999; SB05) a too high value of $Rm$ can be detrimental for
dynamos because of flux expulsion from the convection
eddies. Especially at Prandtl numbers of the order unity or 
somewhat lower a finite amplitude of the magnetic field is required
for dynamos close to their minimum value of $R$ since the velocity
field modified by the action of the Lorentz force is more suitable for
dynamo action,  mainly through the suppression of the differential
rotation (Grote and Busse, 2001), than the non-magnetic velocity
field. In a statistical sense convection-driven dynamos thus usually
correspond to saddle node bifurcations. 

In SB05 it has been noted that a transition from dynamos dominated by
non-axisymmetric components of the magnetic field to those dominated
by the axisymmetric poloidal field occurs with increasing Prandtl
number when a value $P=P_t\sim 8$ is reached in the case
$\tau=5\times10^3$. In figure \ref{f.0080} results are displayed for the case
$\tau=10^4$ which also exhibit this transition. 
The magnetic energies $\overline{M}_p$, $\overline{M}_t$,
$\check{M}_p$, $\check{M}_t$ are defined in  analogy to the
definitions (3) for the kinetic energies with $h$ and $g$  
replacing $v$ and $w$. For $P<P_t$,  $\overline{M}_p<\check{M}_p$ holds and for $P>P_t$
the opposite is true. The results of figure \ref{f.0080} demonstrate that the 
transition Prandtl number $P_t$ depends strongly on $P_m$ with higher 
values of $P_m$ favoring higher values of $P_t$.
For this reason a more complete
analysis of the transition in the case $\tau=5\times10^3$ has been made
with the inclusion of the Rayleigh number. 
The results shown in figure \ref{f.0090} indicate that the parameter $P_m$ appears 
to have a lesser influence on $P_t$ at $\tau= 5\times10^3$ than at 
$\tau=10^4$. The trend that with increasing $\tau$ the value of $P_t$
decreases is confirmed in figure \ref{f.0090.1} where results obtained for
$\tau=3\times10^4$ and for $\tau=10^4$ are compared. In each of the two cases the Rayleigh
number $R$ has been kept constant. 
While the energies of the axisymmetric components of the magnetic field increase relative to
those of the non-axisymmetric components with increasing $P$,
only a weak tendency in this direction can be noticed in the
corresponding Ohmic dissipations.  

In SB05 the increase in the efficiency of convection in the presence
of a dynamo generated magnetic field has been demonstrated for Prandtl
numbers of the order unity or less. At higher Prandtl numbers the
magnetic field is no longer needed to oppose the destructive (for the
convective heat transport) effects of the differential rotation which
manifest themselves, for instance, in the relaxation oscillations
shown in figure \ref{f.0020}. When Nusselt numbers and kinetic energies are thus
compared for $P=20$ in figure \ref{f.0100} for  convection with and without a
dynamo, little change is seen. The energy of differential rotation even
appears to be slightly larger in the magnetic case.

\section{Oscillatory dynamos}

It has long been known that mean-field dynamos in the presence of a strong
differential rotation typically occur in the form of oscillatory
dynamos, especially if any meridional circulation is relatively weak
(Roberts, 1972). In the present case of dynamos driven by turbulent
convection nearly time periodic oscillations are a persistent
feature in the case of hemispherical and quadrupolar dynamos. 
Typical examples have been presented by Grote et al. (1999) and Grote
and Busse (2000). Oscillatory dynamos with dipolar symmetry are less
frequently encountered. An example is shown in figure
\ref{f.0130}. Since the prograde differential rotation increases with 
radius $r$ the oscillations assume the form of two symmetric dynamo
waves propagating from the equator to high latitudes where they
decay similarly as found in mean-field dynamos. 

For the theoretical description of these waves we follow Parker's
(1955) plane layer model with the $(x,y,z)$-coordinates corresponding
to the $(\varphi,-\theta,r)$-directions in the spherical shell. With the
ansatz 
\begin{equation}
\bec B = \bec B_p + \vec i B, \quad \bec B_p = \nabla \times \vec i A, \quad\vec v = \vec i \,U + \check{\vec v},
\end{equation}
where $\vec i$ is the unit vector in the $x$-direction and where the
fields $A, B$ are $x$-independent, $U$ depends only on $z$ and the
$x,y$-average of $\check{\vec v}$ vanishes. Using Parker's equations
in dimensionless form, 
\begin{equation}
\frac{\partial}{\partial t} A =\hat \alpha B + \nabla^2A/P_m, \quad \frac{\partial}{\partial t} B = \bec B_p\cdot \nabla U +  \nabla^2 B/P_m, 
\end{equation}
we may obtain solutions of the form
\begin{equation}
(A, B) = (\hat A, \hat B)\, \exp[i\vec q\cdot \vec x + \sigma t] .
\end{equation}
The equations are satisfied with
\begin{equation}
p\hat A = \hat\alpha \hat B, \quad p\hat B = - i (\vec q \times \nabla U)_x \hat A \:\mbox{ with}\quad p = \sigma + |\vec q|^2/P_m
\end{equation}
and with the resulting dispersion relation
\begin{equation}
p^2 = 2 i \Gamma \equiv 2 i (-\hat \alpha (\vec q \times \nabla U)_x /2).
\end{equation}
The latter yields the growth rates
\begin{equation}
\sigma = \left\{\begin{array}{lr}
    - |\vec q|^2/P_m \pm \sqrt{\Gamma} \pm i \sqrt{\Gamma} \qquad
    \mbox{for}\quad \Gamma > 0, \\[1ex]
    -  |\vec q|^2/P_m \pm \sqrt{|\Gamma|} \mp i \sqrt{|\Gamma|}\qquad
    \mbox{for} \quad \Gamma < 0.
  \end{array}\right.
\end{equation}
Growing waves are found for $\mp\hat \alpha q_y dU/dz > 2 |\vec
q|^4/P_m^2$ with the wave propagating in the $\mp y$-direction
provided that $q_y$ and $dU/dz$ are positive quantities. Here the
upper (lower) sign refers to the case $\Gamma>0$ ($\Gamma<0$). 
\myfigure{
\begin{center}
\epsfig{file=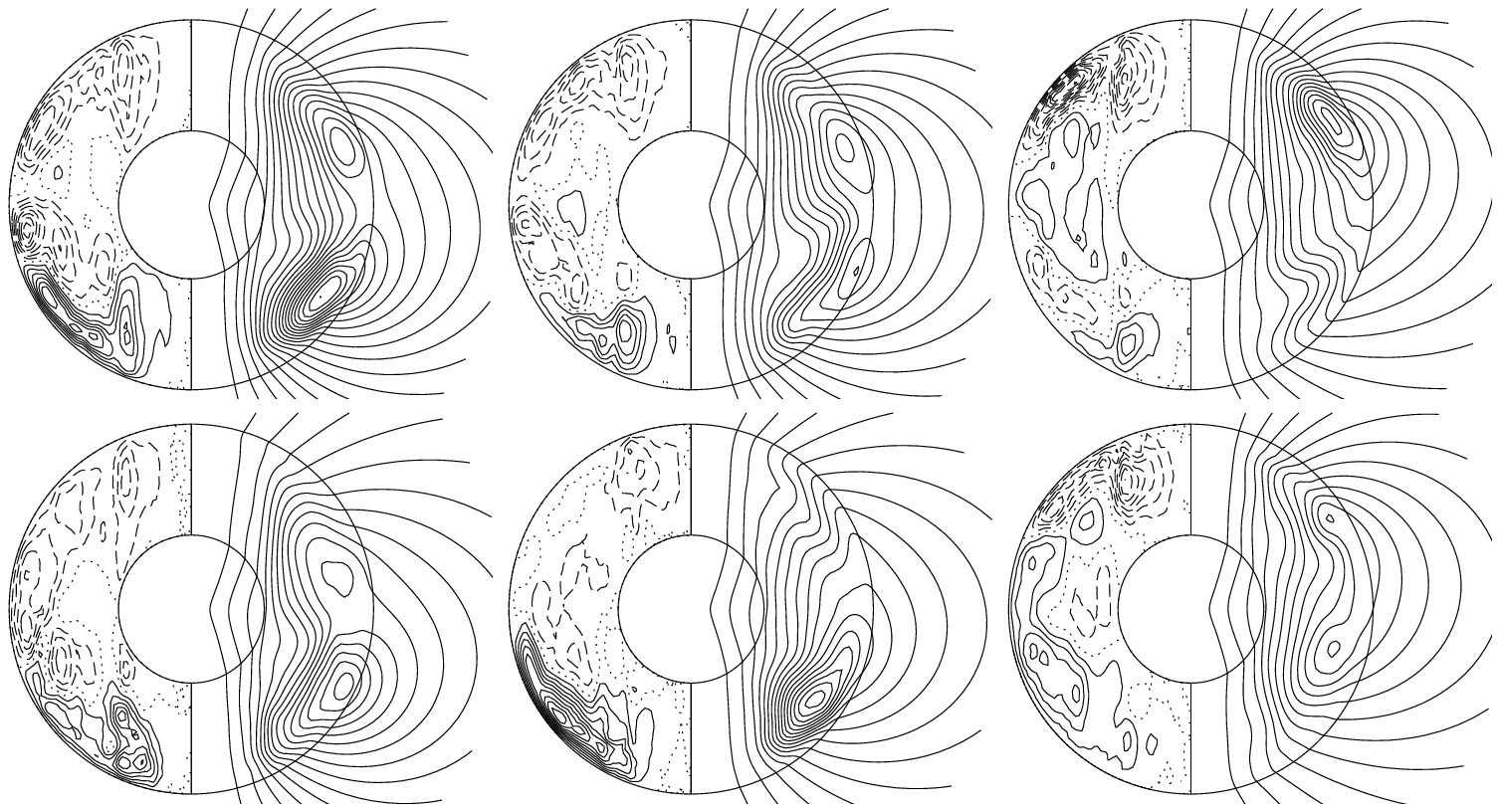,width=11cm}
\end{center}
}{
  Same as figure \ref{f.0130} but for $P=0.05$,  $\tau=10^5$,
  $R=4 \times 10^6$, $P_m=0.3$ with $\Lambda=1.1364$. with $\Delta t = 0.0035$.
  The time series of plots starts with the upper left plot and continues
  in the clockwise sense.
}{f.0160}

The coefficient $\hat\alpha$ is related to the helicity of the convection flow,
\begin{equation}
\hspace{-1.5cm}\hat\alpha \equiv - \frac{1}{3P_m}\int \int \frac{\hat q^2 F(\hat q,\omega)}{\omega^2 + \hat q^4/P_m^2}d\hat q d\omega \approx - \frac{P_m}{3\hat q^2}\int \int F(\hat q,\omega)d\hat q d\omega  \equiv  - \frac{P_m}{3\hat q^2} \langle \check{\vec v} \cdot \nabla \times \check{\vec v} \rangle.
\end{equation}
where $F(\hat q,\omega)$ is the helicity spectrum function (Moffatt,
1978) which is assumed to have a peak at the typical wavenumber $\hat
q$ with $\hat q^2/P_m \gg |\omega|$. $\hat q$ will typically exceed
the value $|\vec q|$ used in expression (5.3). But if we use an
average value $q_y \approx \hat q \approx 2\pi/d$, we find for the
dimensionless frequency $\omega$, 
\begin{equation}
\omega \approx \frac{1}{2\pi}\left(P_m\frac{\pi}{3} \langle \check{\vec v} \cdot \nabla \times \check{\vec v} \rangle\sqrt{2\overline{E_t}}\right)^{1/2}.
\end{equation}
Periods corresponding to this value of $\omega$ are presented for
several dipolar, hemispherical and quadrupolar dynamos in figure \ref{f.0120}
where they can be compared with periods obtained from the  
dynamo simulations. For this purpose the helicity $\langle \check{\vec
  v} \cdot \nabla \times \check{\vec v}\rangle$ was obtained as an
average over time and over either the northern or the southern
hemisphere. Even though the expression (5.8) represents only a crude
estimate it seems to fit the various oscillatory dynamos quite well.
The agreement between mean-field theory and numerical simulation is
due to the dominant action of the differential rotation in creating the
axisymmetric toroidal magnetic flux. As discussed in more detail in
SB05 the $\alpha$-effect of mean-field theory is usually not clearly
noticeable in numerical simulations in contrast to the $\omega$-effect
(see also Schrinner \etal.~2005)  

Because the Prandtl number is moderately large in the case of figure \ref{f.0130} the typical polar
flux tubes have formed which do not participate in the
oscillation. The axisymmetric poloidal field, however, participates
fully in the oscillation and exhibits about equal amplitudes for both
polarities. This property is often missing as shown in figure \ref{f.0140} where
the red field lines dominate and a dipole field with the opposite
polarity does not become established throughout the cycle. Finally, an
extreme version of the latter situation is found in the case of the so-called
"invisibly" oscillating dynamos (Busse and Simitev, 2005). A typical example is found when the magnetic Prandtl number is
somewhat lower than in the case of figure \ref{f.0130} such that the polar flux tubes are
much stronger and the oscillation of the
axisymmetric poloidal field is suppressed. As a result
external observers will not be able to recognize the oscillation of
the axisymmetric toroidal field outside the tangent cylinder. 

Besides the oscillations related to traveling dynamo waves other types
of oscillations may be noticed. Typical for Prandtl numbers of the
order unity or less are global oscillations in which a nearly
hemispherical field switches periodically the hemispheres as shown in
figure \ref{f.0160}. This behavior reminds one of similar, but much weaker
oscillations with a period of about 3.6 years which have been reported
for the solar magnetic field (Knaack \etal \,2004). According to the
observations the flux emerging from the solar southern hemisphere
periodically exceeds that from the northern hemisphere and vice versa.
\myfigure{
\begin{center}
\epsfig{file=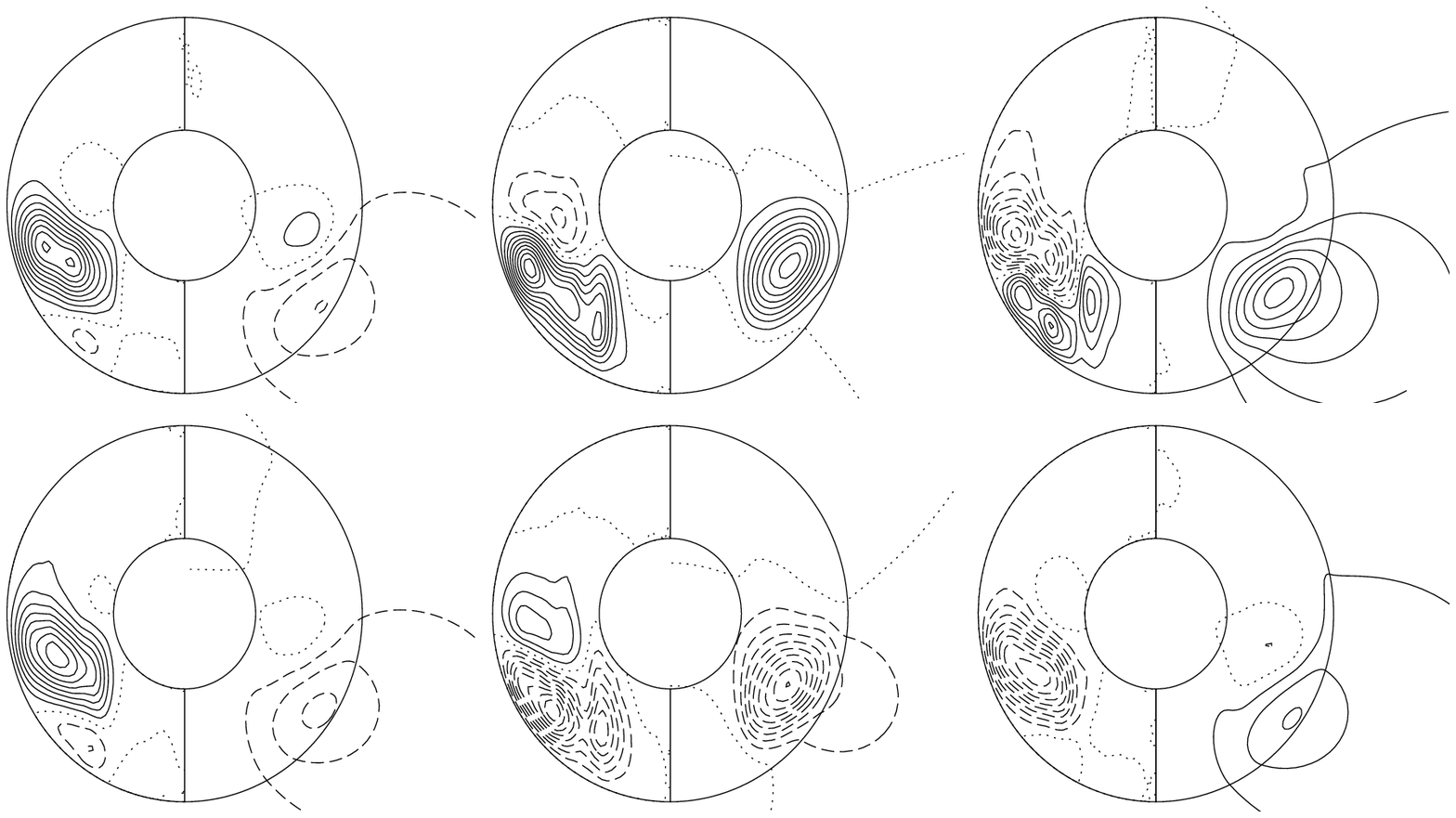,width=11cm,height=6cm}
\end{center}
}{
A period of hemispherical oscillations for $P=0.1$,  $\tau=10^5$,
$R=6 \times 10^6$, $P_m=0.11$ with $\Lambda=0.1424$. 
The plots follow clockwise with $\Delta t =0.005$ and show isolines
similar to those in figure \ref{f.0130}. 
}{f.0170}

\myfigure{
\mypsfrag{Md}{$M_{dip}$}
\mypsfrag{Mq}{$M_{quad}$}
\mypsfrag{Es}{$E$}
\mypsfrag{0}{0}
\mypsfrag{5}{5}
\mypsfrag{10}{10}
\mypsfrag{1}{1}
\mypsfrag{2}{2}
\npsfrag{t}{tl}{$t$}
\mypsfrag{x}{\hspace*{-4mm}$\times10^4$}
\npsfrag{0.05}{tl}{0.05}
\npsfrag{0.07}{tl}{0.07}
\npsfrag{0.09}{tl}{0.09}
\npsfrag{0.11}{tl}{0.11}
\npsfrag{0.13}{tl}{0.13}
\begin{center}
\hspace*{-8mm}
\epsfig{file=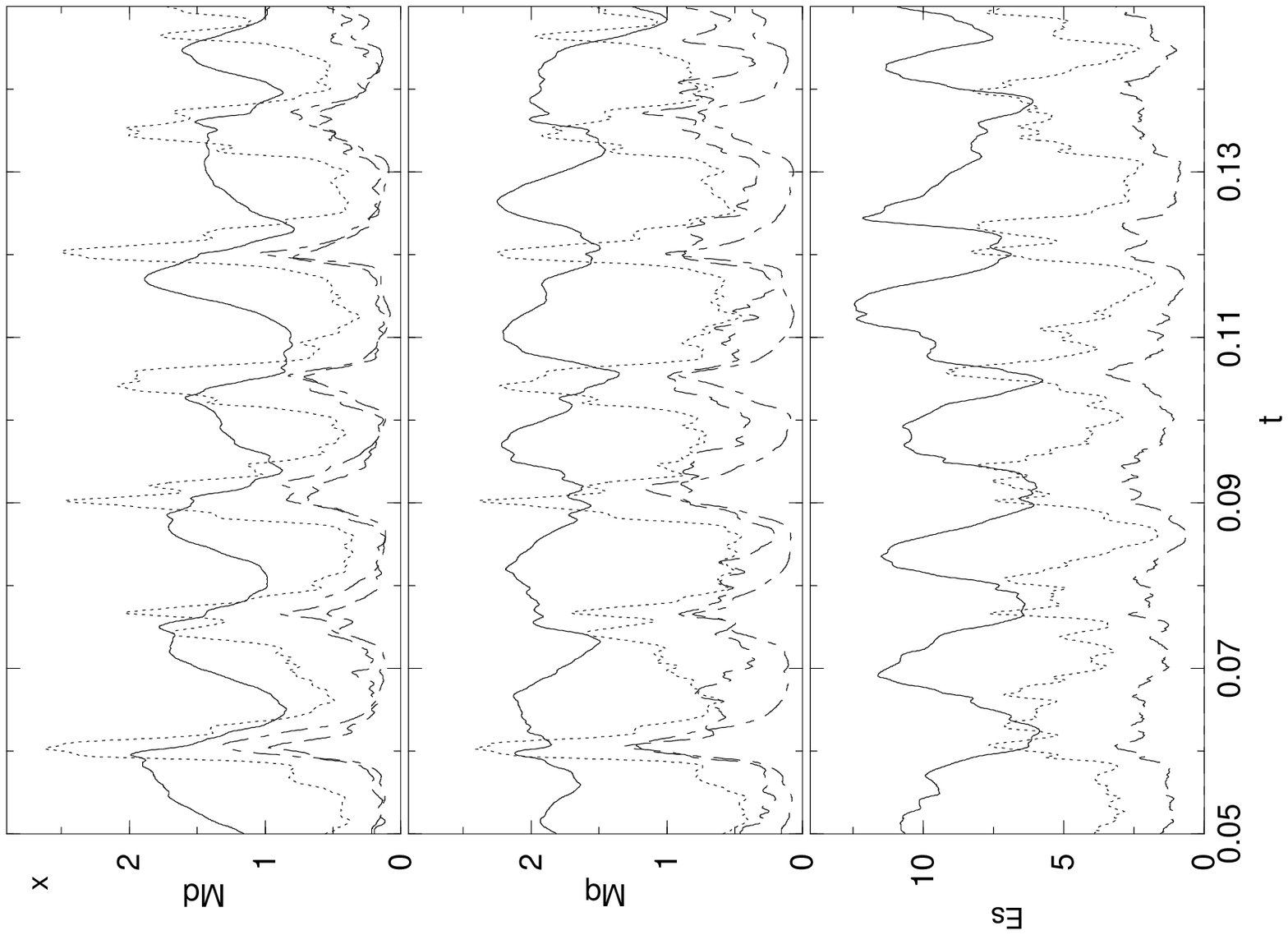,height=8cm,width=10cm,angle=-90}
\end{center}
}{
Time series of energy densities of a hemispherical dynamo
  in the case $P=0.1$, $\tau=10^5$, $R=6\times10^6$, $P_m=0.11$.
  The upper and middle panels show energy densities of
  dipolar and quadrupolar components of the magnetic field, while the
  lower panel shows energy densities of the velocity field.
  The mean toroidal components are represented by solid lines, the
  fluctuating toroidal - by dotted lines, the mean poloidal - by
  dot-dashed lines and the fluctuating poloidal by dashed lines.
}{f.0180}

While the dynamo oscillations can  be well understood in terms of the
linear Parker model, they actually occur at finite amplitudes of the
magnetic field. As a measure of the latter we have denoted the
Elsasser number $\Lambda$ in the captions of figures  \ref{f.0130},
\ref{f.0140}, \ref{f.0160} and \ref{f.0170}. The Elsasser number is
defined by  
\begin{equation}
\Lambda= \frac{2 M P_m}{\tau}
\label{e.1000}
\end{equation}
where $M$ denotes the magnetic energy density
averaged over the fluid shell and in time, \ie
$M=\overline{M}_p+\overline{M}_t+\check{M}_p+\check{M}_t$. The
interaction between magnetic and velocity field has been taken into
account in expression (5.8), of course, in that the actual values for
$\overline{E_t}$ and the helicity have been used in its evaluation. An
even stronger interaction with the velocity field is found in the case
of figure  \ref{f.0170} where a coupling with relaxation oscillations
of convection similar to those shown in figure \ref{f.0020}
occurs. The period, however, is still comparable 
to that of the dynamo waves discussed above. These oscillations are
typical for hemispherical dynamos of the kind shown in figure
\ref{f.0170} as is evident from figure \ref{f.0120}(a). Hardly a full
wavelength of the standing dynamo wave participates in the
oscillation in this case and thus the amplitude of the mean poloidal
field varies throughout the cycle. Accordingly the Lorentz force
restraining the differential rotation varies as well. The
differential rotation thus participates in the oscillation and the
other components of the velocity field affected by its shearing action
do so as well as can clearly be seen in the time record of the kinetic
and magnetic energies displayed in figure \ref{f.0180} for the same case as
shown in figure \ref{f.0170}.  
\myfigure{
\begin{center}
\hspace*{-10mm}
\epsfig{file=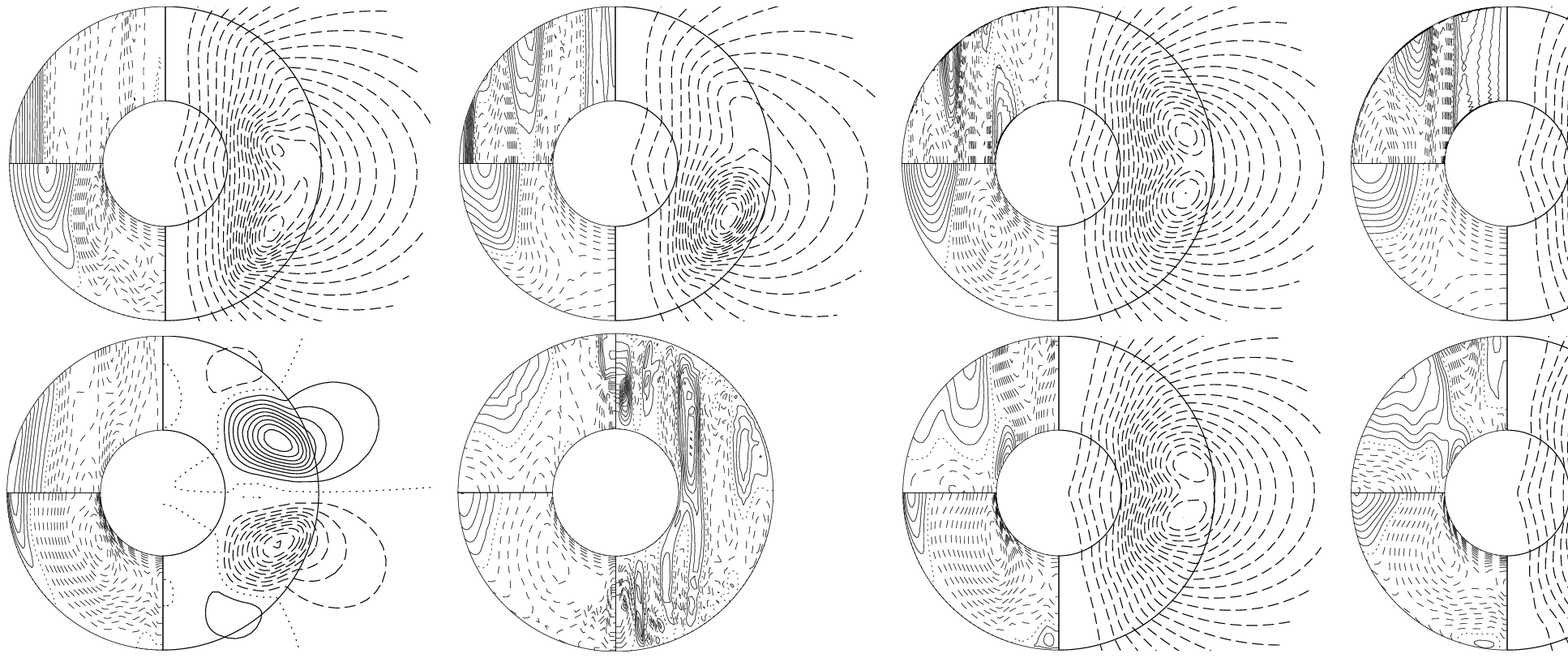,width=11.5cm}
\end{center}
}{
The dynamo cases (a) to (h) presented in table \ref{t.0020} (from left
to right, first row then second row). In each plot the upper left
quarter shows the differential rotation $\overline{u}_\varphi$, the
lower left quarter shows the temperature perturbation
$\overline{\Theta}$ and the right half shows the meridional field lines
$r \sin \theta \dd_\theta \overline{h}=$const. The plot of case (f) shows
the meridional streamlines $r\sin\theta \partial_\theta
\overline{v}=$const.~to the right. 
}{f.0190}

\section{Dynamos for different boundary conditions}

All dynamos discussed so far in the preceding sections have been
computed for stress-free boundaries with fixed temperatures. In this
section some convection-driven dynamos obtained for other boundary
conditions are considered. They have been listed in table \ref{t.0020}. As must
be expected on the basis of the results shown in figure \ref{f.0060} the strength
of convection with fixed heat flux outer boundary is lower than in the
case of the fixed temperature condition. This property is reflected in
the corresponding dynamos where the kinetic and magnetic energies in
cases B are usually less than those in cases A except for the
differential rotation and the associated mean toroidal magnetic
field. 
\renewcommand{\x}{\times}
\begin{table}
\scriptsize
\begin{tabular}{ccccc@{\extracolsep{1mm}}cc@{\extracolsep{1mm}}cc}
Case & (a) & (b) & (c) & (d) & (e) & (f) & (g) & (h) \\
& & & &  & & & &  \\
& 
\multicolumn{4}{c}{$P=0.1$, $\tau=10^5$, $R=4\x10^6$, $P_m=0.5$} &
\multicolumn{2}{c}{$P=5$, $\tau=5\x10^3$}  &
\multicolumn{2}{c}{$P=20$, $\tau=10^4$} \\
&
\multicolumn{4}{c}{} &
\multicolumn{2}{c}{$R=8\x10^5$, $P_m=3$}  &
\multicolumn{2}{c}{$R=2\x10^6$, $P_m=20$} \rule{0in}{3mm}\\  \cline{2-5} \cline{6-7}  \cline{8-9} 
& & & &  & & & &  \\                                                                
BC & {\bf A} & {\bf B} & {\bf C} & {\bf D} & {\bf A} & {\bf  B} &{\bf  A}& {\bf B}  \\
& & & &  & & & &  \\
Type & D & D & D & D & Q & -- & D & D \\                                                                
& & & &  & & & &  \\                                                                
  $\overline{E}_p $ & 0.286$\x10^2$  &  0.258$\x10^2$ &  0.112$\x10^2$ &  0.530$\x10^1$  &  0.157           &  0.166 &  0.018 &  0.026                 \\
 &  & & (0.937$\x10^2$) &  &   (0.147) & &  (0.012) & (0.017) \\
  $\overline{E}_t $ & 0.599$\x10^4$  &  0.641$\x10^4$ &  0.647$\x10^3$ &  0.950$\x10^3$  &  0.533$\x10^2$   &  0.176$\x10^2$     &  0.114$\x10^1$ &  0.166$\x10^1$\\
 &  & & (0.758$\x10^5$) &  &    (0.113$\x10^3$)  & &  (0.141$\x10^1$)& (0.140$\x10^1$) \\
  $\check{E}_p    $ & 0.142$\x10^5$  &  0.109$\x10^5$ &  0.115$\x10^5$ &  0.576$\x10^4$  &  0.574$\x10^2$   &  0.393$\x10^2$     &  0.500$\x10^1$ &  0.361$\x10^1$\\
 &  & & (0.221$\x10^5$) &  &  (0.545$\x10^2$)  & &  (0.758$\x10^1$) & (0.433$\x10^1$)  \\
  $\check{E}_t $    & 0.336$\x10^5$  &  0.273$\x10^5$ &  0.257$\x10^5$ &  0.139$\x10^5$  &  0.119$\x10^3$   &  0.867$\x10^2$     &  0.985$\x10^1$ &  0.777$\x10^1$\\
 &  & & (0.540$\x10^5$) &  &  (0.117$\x10^3$)  & & (0.149$\x10^2$)  & (0.952$\x10^1$)\\
& & & &  & & & &  \\                
  $\overline{M}_p $&  0.129$\x10^5$  &  0.109$\x10^5$ &  0.392$\x10^5$ &  0.305$\x10^5$  &  0.487$\x10^1$   & -- &  0.886$\x10^2$  &  0.104$\x10^3$ \\
  $\overline{M}_t $&  0.133$\x10^5$  &  0.102$\x10^5$ &  0.879$\x10^4$ &  0.104$\x10^5$  &  0.325$\x10^1$   & -- &  0.184$\x10^2$  &  0.260$\x10^2$ \\
  $\check{M}_p$    &  0.136$\x10^5$  &  0.100$\x10^5$ &  0.169$\x10^5$ &  0.112$\x10^5$  &  0.818$\x10^1$   & -- &  0.702$\x10^2$  &  0.723$\x10^2$ \\
  $\check{M}_t$    &  0.307$\x10^5$  &  0.264$\x10^5$ &  0.345$\x10^5$ &  0.257$\x10^5$  &  0.109$\x10^2$   & -- &  0.865$\x10^2$  & 0.871$\x10^2$  \\
& & & &  & & & &  \\                                                               
  $\overline{V}_p $   &  0.643$\x10^5$   &  0.529$\x10^5$     & 0.104$\x10^6$    &  0.795$\x10^5$     &  0.151$\x10^3$   &  0.183$\x10^3$    &  0.152$\x10^2$   &  0.410$\x10^2$ \\
 &  & & (0.108$\x10^8$)  &  &      (0.140$\x10^3$) & &      (0.165$\x10^2$)  &    (0.202$\x10^2$)   \\ 
  $\overline{V}_t $   &  0.584$\x10^6$   &  0.608$\x10^6$     & 0.366$\x10^6$    &  0.295$\x10^6$     &  0.182$\x10^4$   &  0.936$\x10^3$    &  0.114$\x10^3$   &  0.142$\x10^3$ \\
 &  & & (0.280$\x10^8$)  &  &      (0.324$\x10^4$)  & &     (0.134$\x10^3$)  &    (0.179$\x10^3$)   \\ 
  $\check{V}_p    $   &  0.133$\x10^8$   &  0.956$\x10^7$     & 0.179$\x10^8$    &  0.890$\x10^7$     &  0.414$\x10^5$   &  0.311$\x10^5$    &  0.493$\x10^4$   &  0.356$\x10^4$ \\
 &  & & (0.446$\x10^8$)  &  &      (0.390$\x10^5$)  & &     (0.909$\x10^4$)  &    (0.467$\x10^4$)   \\ 
  $\check{V}_t $      &  0.229$\x10^8$   &  0.176$\x10^8$     & 0.354$\x10^8$    &  0.182$\x10^8$     &  0.455$\x10^5$   &  0.378$\x10^5$    &  0.524$\x10^4$   &  0.409$\x10^4$ \\
 &  & & (0.819$\x10^8$)  &  &      (0.445$\x10^5$)  & &     (0.964$\x10^4$)  &    (0.613$\x10^4$)   \\ 
& & & &  & & & &  \\                        
  $\overline{O}_p $   &  0.626$\x10^6$   &  0.490$\x10^6$     & 0.129$\x10^7$    &  0.109$\x10^7$    &  0.544$\x10^3$ & --    &  0.316$\x10^4$   &  0.375$\x10^4$  \\
  $\overline{O}_t $   &  0.148$\x10^7$   &  0.132$\x10^7$     & 0.944$\x10^6$    &  0.858$\x10^6$    &  0.359$\x10^3$ & --    &  0.290$\x10^4$   &  0.438$\x10^4$  \\
  $\check{O}_p    $   &  0.125$\x10^8$   &  0.855$\x10^7$     & 0.138$\x10^8$    &  0.738$\x10^7$    &  0.468$\x10^4$ & --    &  0.519$\x10^5$   &  0.497$\x10^5$  \\
  $\check{O}_t $      &  0.253$\x10^8$   &  0.212$\x10^8$     & 0.268$\x10^8$    &  0.146$\x10^8$    &  0.600$\x10^4$ & --    &  0.558$\x10^5$   &  0.504$\x10^5$   \\
& & & &  & & & &  \\   
  $\Lambda $    & 0.707& 0.576& 0.995& 0.780 &0.032 & --  & 1.055 &  1.160\\
  $Rm $         &  164& 149&  138& 102 & 64 & 50 & 113  &102  \\
  $Nu_i$        &  2.125& 1.926& 2.275 &1.730 & 11.28 & 9.577 & 13.09 & 11.95 \\
 &  & &(2.443)&  &  (10.63)  & &    (13.33) &  (10.44)  \\
  $Nu_o$        &  1.067& 1.055& 1.075 &1.046  & 1.656 &1.423  & 1.753  & 1.566 \\
 &  & &(1.092)&  &  (1.615)  & &    (2.074) &  (1.415) \\    
\end{tabular}
\normalsize
\caption{Time-averaged global properties of dynamos with various
    velocity and thermal boundary conditions as follows. 
    {\bf A}: stress-free and fixed-temperature, 
    {\bf B}: stress-free and fixed-flux,
    {\bf C}: no-slip and fixed-temperature,
    {\bf D}: no-slip and fixed-flux. The
    predominant symmetry type
    is indicated with ``D'' if dipolar, ``Q'' if quadrupolar and ``--'' if
    the dynamo is decaying. Values for non-magnetic convection are
    given in brackets where available.}
\label{t.0020}
\end{table}
\renewcommand{\x}{\cdot}

Of particular interest are the comparisons of cases with and without
magnetic fields. Values for the latter cases have been added in table
1 in brackets wherever they were available. In the cases (a) and (b)
the numerical resolution turned out to be insufficient in the absence
of the magnetic field such that the solution diverges. This result
emphasizes the smoothing effect of the magnetic field in cases with
$P_m < 1$.  At high Prandtl numbers 
the kinetic energies in dynamo cases are always a bit lower than in
the corresponding non-magnetic cases. Only the mean poloidal energy
densities measuring the strength of axisymmetric meridional
circulations and in some instances the mean toroidal energy density
appear to be enhanced by the dynamo action. This result seems to be
independent of the thermal boundary conditions and agrees generally
with the results displayed in figure \ref{f.0100}. This same situation appears to
hold in the case of no-slip conditions at the boundaries even for
Prandtl numbers of the order unity or less as is evident from case (c)
of table 1. This property is in stark contrast to the case of
stress-free boundaries where the energy densities of the
non-axisymmetric components are usually strongly amplified by the
dynamo as is evident from figure 16 and others of SB05. Only the
energy of differential rotation is reduced by the Lorentz force in
both, the no-slip and the stress-free cases. The same observations
apply in the comparison of results obtained by Christensen et
al. (1999) for the two types of boundary conditions. 

The analogy between the results for no-slip boundaries and for
stress-free boundaries at high Prandtl numbers is not surprising if it
is remembered that the presence of Ekman layers at the rigid
boundaries increases strongly the viscous dissipation and thus the
effective Prandtl number relative to the stress-free case. The analogy
holds also for the energy of the axisymmetric dipolar field which
exceeds the energies of all other components of the magnetic field in
the no-slip cases (c) and (d) as well as in the stress-free cases (g)
and (f) at $P=20$ in table 1. 

Characteristic structures of velocity and magnetic fields of the
examples of table 1 are shown in figure \ref{f.0190}. Only a quarter of the
respective circles is used to visualize the differential rotation and
the axisymmetric component of $\Theta$ since these fields are nearly
symmetric relative to the equatorial plane. 
   
\section{Concluding remarks}

It is popular in dynamo theory to neglect the momentum advection terms
in the equation of motion (Glatzmaier and Roberts, 1995a; Jones, 2003)
since they are undoubtedly small in comparison with the Coriolis
force. In view of the results of section 4 and the analysis of SB05
this neglect can be well justified in the limit of high Prandtl
numbers and to some extent in the presence of strong influences of
Ekman layers. It is not surprising that dynamos without momentum
advection terms exhibit the typical features of high Prandtl number
dynamos such as a dominant axial dipolar magnetic field with strong
toroidal flux tubes in the polar regions. But for values of $P$ of the
order unity or lower and stress-free boundary conditions the neglect
of the momentum advection terms is less well satisfied since quite
different types of dynamos with dominant non-axisymmetric components
of the field are found as seen in figure \ref{f.0080} and even more
dramatically in corresponding figures of SB05. The transition to
dynamos with dominating higher multipoles found by Kutzner and
Christensen (2002) at higher Rayleigh numbers in the presence of
no-slip boundaries must probably be attributed also to the momentum
advection terms. 
In spite of their smallness they can drive the geostrophic
differential rotation which is the most easily exited component of
fluid motion. Neither Coriolis force, nor pressure gradient, nor
buoyancy force can drive a geostrophic differential rotation. Only the
Lorentz force enters as a possible competitor. The latter force
usually tends to damp the differential rotation according to Lenz'
rule since the mean toroidal field is generated by it.  It seems that
the separation of dynamos into these two classes will persist when
$\tau$ increases. But further investigations of this point are
desirable. 

The analysis of this paper has focused on cases of moderate Rayleigh
numbers where the effects of the particular choice of the basic
temperature profile are still noticeable. Other authors have employed
temperature profiles without internal heating and Kutzner and
Christensen (2000) have analyzed ways in which different basic
profiles affect properties of convection driven dynamos. Similarly a
fixed radius ratio has been used in the present paper in order to
avoid the discussion of even more parameters. 
While convection driven dynamos in rotating spherical shells with a
radius ratio of 0.4 or lower typically exhibit dipole and quadrupole
components of the magnetic field that are axially oriented, this
situation tends to change as higher values of the radius
ratio are used. Depending on initial conditions Aubert and Wicht (2004)
find in the case of a radius ratio 0.5 dynamos with an equatorial 
dipole or with an axial dipole for identical values of the parameters of
the problem. Clearly, in applications to cases of particular planetary
or stellar magnetic fields parameters that have been fixed in the
present analysis are likely to play an important role.

\end{document}